\documentclass[a4paper,11pt]{article}
\pdfoutput=1 

\usepackage{jcappub} 

\usepackage[T1]{fontenc} 
\usepackage{braket}
\usepackage{xcolor}
\usepackage{tikz}
\usepackage{cancel}

\newcommand{\kk}{\mathbf{k}}
\newcommand{\xx}{\mathbf{x}}
\newcommand{\qq}{\mathbf{q}}
\newcommand{\vg}{``}

\title{Inflationary Fossils beyond perturbation theory}

\author[a,b]{R. Impavido}
\author[c,d,e]{N. Bartolo}

\affiliation[a]{Dipartimento di Fisica e Scienze della Terra, Universit\`a degli Studi di Ferrara, via Giuseppe Saragat 1, 44122 Ferrara, Italy}
\affiliation[b]{INFN, Sezione di Ferrara, via Giuseppe Saragat 1, 44122 Ferrara, Italy}
\affiliation[c]{Dipartimento di Fisica Galileo Galilei, Universit\`a di Padova, I-35131 Padova, Italy}
\affiliation[d]{INFN Sezione di Padova, I-35131 Padova, Italy}
\affiliation[e]{INAF-Osservatorio Astronomico di Padova, Italy}

\emailAdd{mpvrcr@unife.it}
\emailAdd{nicola.bartolo@pd.infn.it}

\abstract{In this work we provide the missing link between two approaches aimed at characterizing the effect of long perturbation modes in Inflation. We consider the \textit{Inflationary Fossils'} approach (\cite{Jeong:2012df} and related works) that characterizes the power-spectrum of the inflaton field in presence of other long and non dynamical \textit{fossil} fields, and a technique, appeared in \cite{Celoria_2021}, that computes, beyond perturbation theory, the power-spectrum of a scalar field in presence of a \textit{large} fluctuation of a second field.
We clarify a few points on the applicability of the non-perturbative technique. We prove in six distinct cases, one involving a violation of the consistency conditions, that the non-perturbative approach, once expanded to first order in the coupling, matches the perturbative result following the Fossils' approach. \newline 
We believe that this non-perturbative technique extends to all orders the Fossils' approach, resumming infinitely many diagrams of standard in-in perturbation theory.}

\begin{document}

\maketitle
\flushbottom

\section{Introduction}
Inflation is the dominant paradigm for answering many questions regarding our Universe and the origin of its structure with a minimal set of assumptions. \cite{Linde:1981mu, Starobinsky:1980te, Guth:1980zm} The best motivated and simplest models of Inflation have been subject to continuous theoretical and experimental study over the last forty years, with the work, among others, of the {\it Planck} Collaboration constraining inflationary parameters from the  large angular scales of the Cosmic Microwave Background (CMB) \cite{Planck:2018jri}, and the recent data analysis from smaller scales of the ACT collaboration (see, e.g. \cite{calabrese2025atacamacosmologytelescopedr6}).
A fundamental and yet to be answered question regarding Inflation is the particle and interaction content that was present during the inflationary era; such question can be rephrased in terms of the primordial non-Gaussianities \cite{Gangui_1994, Acquaviva:2002ud, Bartolo_2004, Seery_2008} of the theory, where computations have been carried out with approaches based on equations of motion within cosmological perturbation theory, or with the in-in formalism \cite{Maldacena_2003,Weinberg_2005, Chen_2017}, based on Quantum Field Theory.  Subsequently, many ideas from the paradigm of Effective Field Theories have been successfully applied to Inflation, starting with \cite{Cheung_2008, Weinberg_2008, Chen_2007}, in the attempt to provide a tool to describe different classes of inflationary models within a unifying framework. Recently there has also been a growing effort to apply quantum field theoretic techniques to Inflation in order to go beyond perturbation theory, both analytically \cite{Riotto_2008, Guilleux:2016oqv, Celoria_2021, Ach_carro_2022, Werth:2023pfl, Pinol:2023oux} and numerically \cite{Fujita_2013, Fujita_2014, Caravano_2024, Caravano_2025, Caravano:2025diq}, motivated by the will to push further our theoretical understanding and by phenomenological reasons, such as the production of Primordial Black Holes by large density fluctuations, which may be too large to be treated perturbatively.  \newline 
On the other hand, in order to inquire the non-Gaussianities present in the Early Universe, a perturbative approach to seek those very long wavelength and ancient (\vg fossil'') fields in the power spectrum of the curvature perturbation was developed \cite{ Jeong:2012df,Dai_2012, Dai:2013ikl, Dimastrogiovanni:2014ina, Dimastrogiovanni:2022afr}. \newline In this work we prove that such an approach, which for convenience we call \textit{Fossils' approach}, and which is a purely perturbative prescription, is amenable to be extended beyond perturbation theory, resumming infinitely many in-in contributions, following the technique originally introduced in \cite{Celoria_2021}. This generalization provides a way to apply the Fossils' approach beyond its original domain of validity. A brief remark on the jargon that we will use in this work follows: we will use \vg non-perturbative'' and \vg beyond perturbation theory'' as synonyms in their \textit{weak} meaning, referring to quantities computed without recurring to a perturbative expansion, but not necessarily in a strongly coupled theory. Moreover, we will prove the matching between the perturbative series and the results beyond perturbation theory expanded at first order; it is clear then that such results, still reaching beyond perturbation theory, are analytical functions in the coupling.  \newline 
The paper is structured as follows: in Section 2 we review and apply the technique to some toy models, in Section 3 we compute the corresponding Fossils' limits and check that the Fossils' result matches the one obtained from the technique beyond perturbation theory, once expanded to first order in the coupling. In Section 4 we apply the technique to a concrete model of Inflation, inspired by the single-field slow-roll Inflation cubic action (see, e.g.~\cite{Maldacena_2003}), and check the correspondence with the Fossils' approach. In Section we 5 apply the technique to an inflationary model violating the consistency conditions \cite{Chen_2007}, and check the correspondence with the Fossils' approach, showing that the technique is independent on the consistency conditions. In Section 6 we clarify some points on the resummation and point out some similarities and differences with other approaches \vg beyond perturbation theory '' in existing literature. Finally in Section 7 we conclude remarking the main results.

\section{Integrating out large fluctuations: toy models}
In this section we are carefully going to integrate out the large fluctuations in three different scenarios of a two-field toy model, and get a prediction for the power spectrum - in the resulting one-field scenario - for the remaining field. We start by reviewing the result for cubic and non-derivative couplings between the two fields, and then we generalize to cubic derivative couplings. As we will show, we intend all the fields to be sharply peaked in momentum space around one single momentum: respectively a small momentum for the long-wavelength field $\chi$ and a large momentum for the short-wavelength field $\sigma$. This choice is made to ease the notation in the convolutions, but the technique and the results would not change considering more modes of $\chi$, all much longer than the modes of the $\sigma$ field, and having $\bar{\chi}$ as the asymptotic late-time value.

\subsection{\texorpdfstring{\boldmath $\lambda \chi \sigma^2\; $}: in de Sitter: review}
This subsection reviews the approach of Sec. 3 of \cite{Celoria_2021} where a long-mode $\chi$ field, interacting non-derivatively with the $\sigma$ field, is integrated out. The resulting effective action will be valid in the limit of very large fluctuations of the $\chi$ field.\footnote{It is worth mentioning that in a subsequent work \cite{creminelli2024nonperturbativewavefunctionuniverseinflation}, the authors considered a model of single field slow-roll inflation with resonant features, in which it is possible to apply techniques beyond perturbation theory - a saddle-point approximation - also for typical values of the curvature perturbation at late times $\bar{\zeta} \sim P_\zeta^{1/2}$.  }We consider exact de-Sitter space in conformal coordinates \begin{equation}
    ds^2 = (\eta H)^{-2}[-d\eta^2+d\xx^2].
\end{equation} Adopting the same notations as \cite{Celoria_2021}, the action reads:
\begin{equation}\label{trilinear action}
    S = \int d\eta d^3\xx \left\{ \frac{1}{2\eta^2 H^2}\left[\sigma'^2-(\partial_i \sigma)^2 + \chi'^2-(\partial_i \chi)^2 \right]-\frac{\lambda}{2\eta^4H^4}H\chi \sigma^2\right\}, 
\end{equation}
 where $d \eta = dt /a(t)$ is the conformal time, and primes denote derivatives with respect to it.

The equations of motion in real space read: \begin{equation}
    \sigma''-\frac{2}{\eta}\sigma' -\partial_i^2\sigma + \frac{\lambda}{\eta^2 H^2}H\chi\sigma =0,
\end{equation}
\begin{equation}
    \chi''-\frac{2}{\eta}\chi' -\partial_i^2\chi + \frac{\lambda}{2\eta^2 H^2}H\sigma^2 =0.
\end{equation}
We define the spatial Fourier transforms of the fields as \begin{equation}
    \sigma(\kk)= \int \frac{d^3\kk}{(2\pi)^3}e^{i \kk \xx}\sigma(\xx).
\end{equation}
We now lay down the main steps of the approach.
\paragraph{First requirement: \boldmath $ \lambda \ll 1 $.} The theory is weakly coupled. In this way the $\chi $ equation is a free wave in de Sitter up to a tiny correction, and reads in momentum space: \begin{equation}
    \chi''(\kk)-\frac{2}{\eta}\chi'(\kk)+ k^2\chi(\kk)=0, 
\end{equation}where $k = |\kk|$. If we suppose that $\chi$ has only one mode in Fourier space, $k_\chi$, this equation has a solution like \begin{equation}
    \chi(\kk,\eta)=\delta^{(3)}(\kk-\kk_\chi)(2\pi)^3\bar{\chi}(1-ik\eta)e^{ik\eta}, 
\end{equation}where a dimensional analysis check tells that $[\bar{\chi}]=1$. 
\paragraph{Second requirement: \boldmath $\bar{\chi}\lambda/H\sim 1$.}
If we imagine that the fluctuation of $\chi$ is much larger than its usual fluctuation (that is, in de Sitter space, of order $H$), an unusual behavior of the perturbation theory emerges. In fact, this enhancement prevents us from getting rid of the interacting term also in the $\sigma$ equation (as we did in the previous paragraph), as it is less suppressed than the corresponding term in the $\chi$ equation. $\sigma$ then is still \vg feeling'' the $\chi$ contribution. This is transparent if we consider the $\sigma$ equation in Fourier space, which reads \begin{align}
     &\sigma''(\qq)-\frac{2}{\eta}\sigma'(\qq)+q^2\sigma(\qq)+\frac{\lambda}{\eta^2H}\int \frac{d^3\kk}{(2\pi)^3}\chi(\qq)\sigma(\qq-\kk) =  \notag\\
    & \sigma''(\qq)-\frac{2}{\eta}\sigma'(\qq)+q^2\sigma(\qq)+\frac{\lambda}{\eta^2H}\bar{\chi}(1-ik_\chi\eta)e^{ik_\chi\eta}\sigma(\qq-\kk_\chi) = 0.
\end{align}
We cannot, in fact, simply dub the last term as \vg subleading'', due to the $\bar{\chi}$ enhancement.  
\paragraph{Third requirement: \boldmath $\kk_\chi \ll \qq$.} If we suppose all the modes of $\sigma$ are much shorter than the mode of $\chi$, then $\sigma(\qq-\kk_\chi)\sim \sigma (\qq) $. We will assume this simplification also in subsequent sections, whenever is needed. 
\paragraph{Fourth requirement: \boldmath $k_\chi \eta \to 0$.} In this time range, hence at late times, the long wavelength $\chi$ oscillation is approaching the asymptotic (constant) limit of a free wave in de Sitter space. We can think of it as \vg frozen'', and we can neglect the oscillating terms. The equation then reads \begin{equation}
    \sigma''(\qq)-\frac{2}{\eta}\sigma'(\qq)+q^2\sigma(\qq)+\frac{\lambda}{\eta^2H}\bar{\chi}\sigma(\qq)=0, 
\end{equation} which is the equation for a massive scalar field in de Sitter, with $m^2=  H\bar{\chi}\lambda$. Given that the $\chi$ field is no longer dynamical, we can substitute it into the action, and get an \vg effective'' action: \begin{equation}
    S_\text{eff}= \int d\eta d^3\xx \left\{-\frac{1}{2}\sqrt{-g}\left(\eta^2 H^2(\sigma'^2 - (\partial_i \sigma)^2)+{\lambda \bar{\chi}H}\sigma^2\right)\right\}, 
\end{equation} which is a functional of the sole $\sigma$ field. The power spectrum, obtainable by solving exactly the equation of motion without having to recur to perturbative techniques (and matching with the Bunch-Davies vacuum in the UV), reads \begin{equation}
    \braket{\sigma_{\kk_1}\sigma_{\kk_2}}_\text{eff}'=\frac{H^2}{2k_1^3}\left(\frac{k_1}{aH}\right)^{3-2\sqrt{9/4-m^2/H^2}}= \frac{H^2}{2k_1^3}\left(-k_1 \eta \right)^{3-2\sqrt{9/4-\lambda \bar{\chi} /H}},
\end{equation}
where ' here and next to power spectra stands for the omission of the $(2\pi)^3\delta^{(3)}(\kk_1+\kk_2)$ factor.
This is one of the results of \cite{Celoria_2021}. This power spectrum, expanded to first order in $\lambda$, reads \begin{equation}\label{np cubic}
    \braket{\sigma_{\kk_1}\sigma_{\kk_2}}_\text{eff}'\sim\frac{H^2}{2k_1^3}\left(1-\frac{2}{3}\text{log}(-k_1\eta)\frac{\lambda \bar{\chi}}{H}\right),
\end{equation} where $\eta$ is intended in the (divergent) limit to 0, at late times when also the mode of $\sigma$ has crossed the horizon.  
\subsection{\boldmath \texorpdfstring{$\lambda \partial_0 \sigma \partial^0\sigma \chi$}: in de Sitter}
In this section and the following we will explore derivative coupling, inspired by $\lambda g^{\mu \nu}\partial_\mu \sigma \partial_\nu \sigma \chi $. We will keep the temporal part and the spatial part separated, to ease the computation and to highlight the differences between the two. The complete case, in a more concrete scenario, will be treated in Section 4. \\ 
We proceed similarly to how we did before. The starting action is \begin{equation}\label{timederivatives}
    S= \int d\eta d^3\xx \left\{-\frac{1}{2}\sqrt{-g}\left[\partial_\mu \sigma \partial_\nu \sigma g^{\mu \nu} + \partial_\mu \chi \partial_\nu \chi g^{\mu \nu} - \lambda \frac{\chi}{H} \partial_0\sigma \partial_0 \sigma g^{00}\right]\right\}. 
\end{equation}
The equation of motion of $\chi$ reads \begin{equation}\label{chi in deriv coupling}
    \chi''-\frac{2}{\eta}\chi'+k^2\chi = -\frac{\lambda}{2H}\int \frac{d^3\qq}{(2\pi)^3}\sigma'(\qq) \sigma'(\kk-\qq).
\end{equation}
The equation of motion for $\sigma$ reads in Fourier space: \begin{align}
    \sigma''(\qq)-\frac{\lambda}{H}\int \frac{d^3\kk}{(2\pi)^3}\sigma''(\kk)\chi(\qq-\kk)-\frac{2}{\eta}\sigma'(\qq)+\frac{2\lambda}{\eta H}\int \frac{d^3\kk}{(2\pi)^3}\sigma'(\qq-\kk)\chi(\kk) \notag \\ -\frac{\lambda}{H}\int \frac{d^3\kk}{(2\pi)^3}\chi'(\kk)\sigma'(\qq-\kk)+ q^2\sigma(\qq) =0. 
\end{align}
Notice that in the derivation of the equation for $\sigma$, as we are in presence of derivative couplings, one must also take into account the terms coming from deriving $\chi$ in the covariant derivative \begin{equation}
    \nabla_\mu \left(\frac{\partial \mathcal{L}}{\partial \partial_\mu \sigma}\right), 
\end{equation}
resulting in the first term of the second line in the e.o.m. 
This system of coupled equations of motion is more complicated than the previous one. One must keep in mind that, albeit we have not yet solved the system and therefore we have not an explicit solution for the modes, a time derivative of an oscillating function will produce a contribution proportional to the momentum of the function: $\sigma'\sim k_\sigma \sigma$. To simplify the system we can explore a parameter space which is essentially three-dimensional, in fact the amplitude of the long field $\bar{\chi}$, the coupling $\lambda$, and the ratio between the two typical momenta $k_\chi /k_\sigma$ are all parameters that from first principles should be independent from one another.  We impose as before the hierarchies $\lambda \ll 1$, $k_\chi /k_\sigma \ll 1$, $\lambda \bar{\chi}/H \sim 1 $, and furthermore we impose \begin{align}
    \lambda \frac{k_\sigma^2}{k_\chi^2}\ll 1. 
\end{align}
This last inequality, in particular, is useful to consider subleading the term on the right hand side of Eq. (\ref{chi in deriv coupling}), making $\chi$ a free field in de Sitter. 
These hierarchies among the parameters makes the equation for $\chi$, after the $\chi$ mode exits the horizon, a free equation, and we can find as a solution the free wave in de Sitter, just as before: \begin{equation}
    \chi(\kk,\eta)=\delta^{(3)}(\kk-\kk_\chi)(2\pi)^3\bar{\chi}(1-ik\eta)e^{ik\eta}. 
\end{equation}
When we insert this solution into the equation for $\sigma$, we can resolve the convolutions and the result reads: \begin{equation}
    \sigma''\left(1-\frac{\lambda\bar{\chi}}{H}(1-ik_\chi\eta)e^{ik_\chi \eta}\right)-\frac{2}{\eta}\sigma'\left(1-\frac{\lambda\bar{\chi}}{H}(1-ik_\chi\eta)e^{ik_\chi \eta}-\frac{\lambda\bar{\chi}}{2H}\eta^2 k_\chi^2e^{ik_\chi\eta}\right)+ q^2 \sigma =0.
\end{equation}
We can appreciate that in the limit in which the $\chi$ field reaches its asymptotic value, meaning after $k_\chi \lesssim 1/\eta$, all the other terms are driven to 0, and the resulting equation for $\sigma$ is  \begin{equation}
    (1-\lambda\bar{\chi}/H)\left(\sigma '' -\frac{2}{\eta}\sigma'\right)+q^2\sigma=0.
\end{equation}
The corresponding effective action reads \begin{equation}
    S_\text{eff}= \int d\eta d^3\xx \left\{\frac{-1}{2\eta^4H^4}\left(\partial_0\sigma\partial_0\sigma g^{00}(1-\lambda\bar{\chi}/{H})+ \partial_i\sigma \partial_j\sigma g^{ij}\right)\right\}.
\end{equation}
We recognize the term $c_s^{-2}=(1-\lambda\bar{\chi}/{H})$ as a modification of the speed of sound of the leftover field $\sigma$, and hence we
     get as a prediction for its power spectrum \begin{equation}\label{np timeder}
    \braket{\sigma_{\kk_1} \sigma_{\kk_2}}_\text{eff}'= \frac{H^2}{2k_1^3c_s} =\frac{H^2}{2k_1^3}\sqrt{1-\lambda\bar{\chi}/{H}}. 
\end{equation} 
If we expand at first order in $\lambda$ the square root, we get: 
\begin{equation}
    \braket{\sigma_{\kk_1} \sigma_{\kk_2}}_\text{eff} \sim\frac{H^2}{2k_1^3}\left(1- \frac{\lambda\bar{\chi}}{2H}\right).
\end{equation}
\subsection{\boldmath \texorpdfstring{$\lambda \partial_i \sigma \partial^i\sigma \chi$}: in de Sitter}
We conclude the exploration of pure toy models with the spatial derivative cubic interaction. The action is \begin{equation}
    S= \int d\eta d^3\xx \left\{-\frac{1}{2}\sqrt{-g}\left[\partial_\mu \sigma \partial_\nu \sigma g^{\mu \nu} + \partial_\mu \chi \partial_\nu \chi g^{\mu \nu} + \lambda \frac{\chi}{H} \partial_i\sigma \partial_j \sigma g^{ij}\right]\right\}. 
\end{equation}
The two equations of motion read, beginning with the one for $\chi$: \begin{equation}
    \chi''-\frac{2}{\eta}\chi'+k^2\chi=-\frac{\lambda}{H}\int \frac{d^3\kk}{(2\pi)^3}\kk \cdot (\qq-\kk)\sigma(\kk)\sigma(\qq-\kk),
\end{equation}
and the equation for $\sigma$ is \begin{equation}
    \sigma''-\frac{2}{\eta}\sigma'+k^2\sigma + \frac{\lambda}{H}\int \frac{d^3\qq}{(2\pi)^3}(\kk-\qq)\cdot (\kk-\qq)\sigma(\kk-\qq)\chi(\qq)-\frac{\lambda}{H}\int \frac{d^3\qq}{(2\pi)^3}\qq\cdot (\kk-\qq)\chi(\qq)\sigma(\kk-\qq)=0. 
\end{equation}
In this case we can see how the spatial derivative acts adding a factor of momentum in front of the object it derives. 
To simplify the system we again have freedom to choose three parameters: the amplitude of the long field $\bar{\chi}$, the coupling $\lambda$, and the ratio between the two typical momenta $k_\chi /k_\sigma$. We impose as before the hierarchies $\lambda \ll 1$, $k_\chi /k_\sigma \ll 1$, $\lambda \bar{\chi}/H \sim 1 $, and as the previous case we impose \begin{align}
    \lambda \frac{k_\sigma^2}{k_\chi^2}\ll 1. 
\end{align}
Again this hierarchy is enough to make the equation for $\chi$, after the $\chi$ mode exits the horizon, a free equation, and we can impose the usual solution: \begin{equation}
    \chi(\kk,\eta)=\delta^{(3)}(\kk-\kk_\chi)(2\pi)^3\bar{\chi}(1-ik\eta)e^{ik\eta}. 
\end{equation}
This insertion makes the $\sigma$ equation: \begin{equation}
    \sigma''-\frac{2}{\eta}\sigma'+k^2\sigma + \frac{\lambda\bar{\chi}}{H}k^2\sigma(1-ik_\chi\eta)e^{ik_\chi\eta}-\frac{\lambda \bar{\chi}}{H} \kk \cdot \kk_\chi \sigma (1-ik_\chi\eta)e^{ik_\chi\eta}=0.
\end{equation}
In the limit in which we neglect $k_\chi $ with respect to $k_\sigma$, we have  
\begin{equation}
    \sigma''-\frac{2}{\eta}\sigma'+ \left(1+\frac{\lambda\bar{\chi}}{H}\right)k^2\sigma=0,
\end{equation}
The corresponding effective action reads
\begin{equation}
    S_\text{eff}= \int d\eta d^3\xx \left\{-\frac{1}{2}\sqrt{-g}\left(\partial_0 \sigma \partial_0 \sigma g^{00}+\partial_i \sigma \partial_j \sigma g^{ij}\left(1+\lambda\bar{\chi}/H\right)\right)\right\}.
\end{equation}
We remark that this effective action gives to the $\sigma$ field a non canonical kinetic term. To be as clear as possible we proceed as follows: we rescale the field $\sigma$ by $\phi =\sigma \sqrt{1+\lambda\bar{\chi}/H}$, 
compute the power spectrum for $\phi$ and then translate back to $\sigma$. The effective action for $\phi$ reads \begin{equation}
    S_\text{eff}= \int d\eta d^3\xx \left\{-\frac{1}{2}\sqrt{-g}\left(\partial_0 \phi \partial_0 \phi g^{00}(1+\lambda\bar{\chi}/H)^{-1}+\partial_i \phi \partial_j \phi g^{ij}\right)\right\}, 
\end{equation}
and we can recognize the speed of sound $c_s^{2}=(1+\lambda\bar{\chi}/H)$ (notice that it is the inverse of what it was in the scenario of the previous subsection).  This time $\phi$ is canonical and we can extract the power spectrum without resorting to any perturbation theory: \begin{equation}
    \braket{\phi_{\kk_1}\phi_{\kk_2}}_\text{eff}'=\frac{H^2}{2k_1^3\sqrt{1+\lambda\bar{\chi}/H}}=(1+\lambda\bar{\chi}/H)\braket{\sigma_{\kk_1} \sigma_{\kk_2}}_\text{eff}'. 
\end{equation} Once this result is obtained, we can also smoothly get the power spectrum of interest for us, hence the one of $\sigma$: \begin{equation}\label{np spaceder}
    \braket{\sigma_{\kk_1} \sigma_{\kk_2} }_\text{eff}'= \frac{H^2}{2k_1^3\sqrt{(1+\lambda\bar{\chi}/H})^3}\end{equation}
    Carrying out the expansion in $\lambda$, the first term reads
    \begin{equation}
          \braket{\sigma_{\kk_1} \sigma_{\kk_2} }_\text{eff}'\sim\frac{H^2}{2k_1^3}\left(1-\frac{3 \lambda\bar{\chi}}{2H}\right)\, .
\end{equation}
\section{Inflationary Fossils}
In this section we compute the bispectra for all the toy models we considered above, and employ them in their squeezed limit to infer the correction in the series expansion with respect to $\lambda$ for the power spectra, following the prescription of \cite{Dai:2013ikl, Jeong:2012df, Dimastrogiovanni:2014ina} and related works (in particular, for a more formal and recent treatment, we considered \cite{Dimastrogiovanni:2022afr}\footnote{We also remark that the cases treated here are cubic coupled theories, and we do not consider the case of linear mixing, which has been done - non perturbatively - in the Appendix of \cite{Bordin_2018}. We will comment on the comparison with existing literature in the dedicated subsection.}). \newline 
We consider $\chi$ and $\sigma$, which should remind us the fields we introduced in the previous Section, to be canonically normalized massless fields in de Sitter, hence each with power spectrum \begin{equation}
    \braket{\sigma_{\kk_1} \sigma_{\kk_2}}'=\braket{\chi_{\kk_1} \chi_{\kk_2}}'= \frac{H^2}{2k_1^3}. 
\end{equation}The \vg Fossils' prescription'' consists, in case of non-Gaussianities, hence in case of interactions between the fields, in adding to the power-spectrum a contribution equal to the squeezed limit of a bispectrum, properly normalized:  \begin{equation}
    \braket{\sigma_{\kk_1} \sigma_{\kk_2}}|_{\chi_L}= \braket{\sigma_{\kk_1} \sigma_{\kk_2}}+ \lim_{\qq\to 0}\frac{\bar{\chi}\braket{\sigma_{\kk_1} \sigma_{\kk_2}\chi_\qq}}{\braket{\chi_\qq \chi_{-\qq}}}. 
\end{equation}

The squeezed limit of the bispectrum is intended precisely to encode the contribution that arises when - in this case - the $\sigma$ field is interacting with a long $\chi$ mode ( indicated here with $\chi_L$.)
In this section we omit the primes $'$, that were standing for delta functions, to ease the notation. 
\subsection{\boldmath \texorpdfstring{$\lambda  \sigma^2   \chi$}:: bispectrum }
In this subsection and in the following ones, we employ the diagrammatical notations and conventions of \cite{Chen_2017} for the in-in formalism, working in the so-called $+- $ basis of the propagators. The dashed line represents the $\chi$ field, the full line represents the $\sigma$ field, and (being both massless fields in de Sitter at quadratic order in the action) they both have the same bulk-to-boundary propagator 
\begin{equation}\label{gplusminus}
    G_{-+}(k;\eta',\eta)= \frac{H^2}{2k^3}(1+ik\eta)(1-ik\eta')e^{ik\eta'}e^{-ik\eta}. 
\end{equation} 
The bispectrum is diagrammatically (here and in the other sections) : 
\vspace{5pt}
\begin{equation}
\braket{\sigma_{\kk_1}\sigma_{\kk_2}\chi_ {\qq}}
= 
\vcenter{
    \hbox{
\tikzset{every picture/.style={line width=0.75pt}}

\begin{tikzpicture}[x=0.75pt, y=0.75pt, yscale=-1, xscale=1]

\draw (90,20) arc[start angle=0, end angle=180, radius=30] -- cycle;
\draw[line width=1.5] (20,20) -- (100,20);
\draw[dash pattern=on 4.5pt off 4.5pt] (70,20) -- (60,50);
\filldraw[fill=white, draw=black] (60,50) circle (3.67);

\end{tikzpicture}}}
+ 
\vcenter{
    \hbox{
\tikzset{every picture/.style={line width=0.75pt}}

\begin{tikzpicture}[x=0.75pt, y=0.75pt, yscale=-1, xscale=1]

\draw (90,20) arc[start angle=0, end angle=180, radius=30] -- cycle;
\draw[line width=1.5] (20,20) -- (100,20);
\draw[dash pattern=on 4.5pt off 4.5pt] (70,20) -- (60,50);
\filldraw[fill=black, draw=black] (60,50) circle (3.67);

\end{tikzpicture}}}
= 2\text{Re}\left[
\vcenter{
    \hbox{\tikzset{every picture/.style={line width=0.75pt}}

\begin{tikzpicture}[x=0.75pt, y=0.75pt, yscale=-1, xscale=1]

\draw (90,20) arc[start angle=0, end angle=180, radius=30] -- cycle;
\draw[line width=1.5] (20,20) -- (100,20);
\draw[dash pattern=on 4.5pt off 4.5pt] (70,20) -- (60,50);
\filldraw[fill=white, draw=black] (60,50) circle (3.67);

\end{tikzpicture}}}
 \right].
\end{equation}
Dots correspond to interaction vertices, and we remark also that switching all the black dots to white dots in a diagram (and vice versa) corresponds to taking the complex conjugate of the corresponding integral. 
In this subsection we take in consideration a non-derivative coupling, namely: \begin{equation}
    S \supset \int d\eta d^3\xx \frac{-\lambda \sigma^2\chi }{2\eta^4H^3} 
\end{equation}
from Eq. (\ref{trilinear action}) 
The formal expression for the \vg white '' diagram is
\vspace{5pt}
\begin{align}
    \vcenter{
    \hbox{\tikzset{every picture/.style={line width=0.75pt}}
\begin{tikzpicture}[x=0.75pt, y=0.75pt, yscale=-1, xscale=1]
\draw (90,20) arc[start angle=0, end angle=180, radius=30] -- cycle;
\draw[line width=1.5] (20,20) -- (100,20);
\draw[dash pattern=on 4.5pt off 4.5pt] (70,20) -- (60,50);
\filldraw[fill=white, draw=black] (60,50) circle (3.67);
\end{tikzpicture}}} &= i\lambda \int d\eta \frac{1}{\eta^4H^3}G_{-+}(k_1; \eta',\eta)G_{-+}(k_2; \eta',\eta)G_{-+}(q; \eta',\eta)  \notag\\
&= \lambda \frac{H^3}{8k_1^3k_2^3q^3}e^{ik_T\eta'}\left[i\left(1-\eta'^2\sum_{i\neq j}k_ik_j\right)+\left(-k_T\eta'+k_1k_2q\eta'^3\right)\right]\times  \notag \\
& \int d\eta \frac{1}{\eta^4}e^{-ik_T\eta}\left(1+ik_T\eta-\eta^2\sum_{i\neq j}k_ik_j-i\eta^3\right).
\end{align}
where $k_T = k_1+k_2+q$.
For symbolical convenience we call $[fac]$ the prefactor in the first line of the previous equation, and $[int]$ the integral. 
The solution of the integral is straightforward\footnote{We remark that: $\text{log}(-ik\eta)=\text{log}(-k\eta)+\text{log}(i)=\text{log}(-k\eta)-i\pi/2 $. The $i\pi$ part will eventually multiply a term proportional to $\eta'$ and will therefore vanish in the limit $\eta'\to 0^-$.} (we omit the discussion about the tilt of contour integral to match the Bunch-Davies vacuum in the UV, which is well-known in the literature, see e.g. \cite{Weinberg_2005}): 
\begin{align}
    \text{Re}[int]=\frac{\pi}{6}+\frac{e^{-ik_T\eta'}}{3}\left(\frac{1}{\eta'}\sum_{i\neq j}k_ik_j-\frac{1}{\eta'}\sum_ik_i^2-\frac{1}{\eta'^3}\right)\notag \, , \\
    \text{Im}[int]= -\frac{1}{3}\left[[\gamma_E+\text{log}(-k_T\eta')]\sum_{i\neq j}k_i^3+\frac{k_Te^{-ik_T\eta'}}{\eta'^2}\right],
\end{align}
where $\gamma_E \sim 0.577$ is the Euler-Mascheroni constant.
To evaluate the bispectrum, we take 
\begin{align}
    \braket{\sigma_{\kk_1}\sigma_{\kk_2}\chi_\qq}&= \lim_{\eta'\to 0}2\left[\text{Re}[int]\text{Re}[fac]-\text{Im}[int]\text{Im}[fac]\right] \notag\\
    &= \frac{-\lambda H^3}{12k_1^3k_2^3q^3}\left([\gamma_E+\text{log}(-k_T\eta')]\sum_i k_i^3+k_1k_2q-k_T\sum_ik_i^2\right).
\end{align}
We remark that in the summations above, with a slight abuse of notation, we mean $i$ ranging form $1$ to $3$ with $k_3= q$. 
In the limit in which the $\sigma$ mode exits the horizon, hence $\eta'\to 0^-$ the logarithmic term dominates the behavior of the bispectrum, and eventually diverges. Such a leading term reads: \begin{equation}
    \frac{-\lambda H^3}{12k_1^3k_2^3q^3}\text{log}(-k_T\eta')(k_1^3+k_2^3+q^3). 
\end{equation}
\paragraph{Fossils' limit}
We now take the limit in the Fossils' prescription as {$\qq \to 0$} and hence  $k_1, k_2\to k$, $k_T\to 2k_1$: \begin{align}
    \lim_{\qq\to 0}\frac{\bar{\chi}\braket{\sigma_{\kk_1} \sigma_{\kk_2}\chi_\qq}}{\braket{\chi_\qq \chi_{-\qq}}}&=\bar{\chi}\frac{-\lambda H^3}{12k_1^3k_2^3q^3}\text{log}(-k_T\eta')(2k_1^3)\frac{2q^3}{H^2} \notag \\
    &= \frac{H^2}{2k_1^3}\left(-\frac{2\lambda \bar{\chi}}{3H}\text{log}(-k_T\eta')\right).
\end{align}
We remark that in the last step we dropped a $\text{log}(2)$ term as it is negligible in the limit $\eta'\to 0$. \newline We see explicitly for the first time that, remarkably, the leading term in the non-perturbative approach in Eq. (\ref{np cubic}) correctly matches the first order in perturbation theory following the Fossils' prescription, for the corresponding theory.
\subsection{\boldmath \texorpdfstring{$\lambda \partial_0 \sigma \partial^0\sigma \chi$}:: bispectrum }
We consider now the time derivative coupling, namely the interacting term \begin{equation}
    S \supset \int d\eta d^3\xx \frac{-\lambda \chi \sigma ' \sigma '}{2 \eta^2 H^3} 
\end{equation} In the action \ref{timederivatives}. We remark that, following the diagrammatic approach of \cite{Chen_2017}, a time derivative amounts to the derivative of the propagator in the corresponding interaction. 
The \vg white'' contribution is: \vspace{5pt}
\begin{align}
    \vcenter{
    \hbox{\tikzset{every picture/.style={line width=0.75pt}}
\begin{tikzpicture}[x=0.75pt, y=0.75pt, yscale=-1, xscale=1]
\draw (90,20) arc[start angle=0, end angle=180, radius=30] -- cycle;
\draw[line width=1.5] (20,20) -- (100,20);
\draw[dash pattern=on 4.5pt off 4.5pt] (70,20) -- (60,50);
\filldraw[fill=white, draw=black] (60,50) circle (3.67);
\end{tikzpicture}}} &= i\lambda \int d\eta \frac{1}{\eta^2H^3}\partial_\eta(G_{-+}(k_1; \eta',\eta))\partial_\eta(G_{-+}(k_2; \eta',\eta))G_{-+}(q; \eta',\eta)  \notag \\
&= \lambda \frac{H^3}{8k_1^3k_2^3q^3}e^{ik_T\eta'}\left[i\left(1-\eta'^2\sum_{i\neq j}k_ik_j\right)+\left(-k_T\eta'+k_1k_2q\eta'^3\right)\right]\times \notag \\
& \int d\eta e^{-ik_T\eta}\left(k_1^2k_2^2+iqk_1^2k_2^2\eta\right).
\end{align}
Again for convenience we call $[fac]$ the prefactor in the first line of the previous equation, and $[int]$ the integral. 
The solution of the integral is again straightforward: \begin{align}
    \text{Im}[int]= \left(\frac{k_1^2k_2^2}{k_T}+\frac{qk_1^2k_2^2}{k_T^2}\right)e^{-ik_T\eta'} \notag \\
    \text{Re}[int]= -\eta' \frac{q k_1^2k_2^2}{k_T}e^{-ik_T\eta'}. 
\end{align}
To evaluate fully the bispectrum, we take \begin{align}
    \braket{\sigma_{\kk_1}\sigma_{\kk_2}\chi_\qq}&= \lim_{\eta'\to 0}2\left[\text{Re}[int]\text{Re}[fac]-\text{Im}[int]\text{Im}[fac]\right] \notag \\
    &= \frac{-\lambda H^3}{4k_1^3k_2^3q^3}\left(\frac{k_1^2k_2^2}{k_T}+\frac{qk_1^2k_2^2}{k_T^2}\right).
\end{align}
\paragraph{Fossils' limit} 
Evaluating the Fossils' limit, we compute \begin{align}
    \lim_{\qq\to 0}\frac{\bar{\chi}\braket{\sigma_{\kk_1} \sigma_{\kk_2}\chi_\qq}}{\braket{\chi_\qq \chi_{-\qq}}}&=\bar{\chi}\frac{-\lambda H^3}{4k_1^3k_2^3q^3}\left(\frac{k_1^2k_2^2}{k_T}+\frac{qk_1^2k_2^2}{k_T^2}\right) \frac{2q^3}{H^2} \notag \\
    &= \frac{H^2}{2k_1^3}\left(-\frac{\lambda\bar{\chi}}{2H}\right)
\end{align}
as ${\bf q} \to 0$ and hence $k_1, k_2\to k$, $k_T\to 2k_1$. This time we have no tree-level divergent result, due to the derivative nature of the interaction. We can see then that the Fossils' approach leads to the following modification of the power spectrum: \begin{equation}
    \braket{\sigma_{\kk_1} \sigma_{\kk_2}}|_{\chi_L}= \frac{H^2}{2k_1^3}\left(1-\frac{\lambda\bar{\chi}}{2H}\right).
\end{equation}
Again we notice how this stands in perfect agreement with the result obtained non-perturbatively once expanded to first order, namely our Eq. (\ref{np timeder}). 
\subsection{\boldmath \texorpdfstring{$\lambda \partial_i\sigma \partial^i\sigma \chi$}:: bispectrum}
We are dealing with a spatial derivative coupling, namely the interaction reads:
\begin{equation}
    S \supset \int d\eta d^3\xx \frac{-\lambda \chi (\partial_i\sigma)^2 }{2 \eta^2 H^3}. 
\end{equation}
The presence of the derivative coupling translates in the appearance of a $i\kk$ term for every derivative in the Feynman rules, that justifies the minus sign in front of the coupling, and no other effect is present. We can directly get a $i\kk$, without having to carry on the derivative explicitly in the $G_{-+}(k;\eta',\eta)$ as we did in the case of the time derivative coupling, due to the fact that in de Sitter space we Fourier transform only the spatial coordinates. The \vg white '' contribution to the bispectrum reads, again meaning in the summations $i$ ranging form $1$ to $3$ with $k_3= q$. : 
\vspace{5pt}
\begin{align}
    \vcenter{
    \hbox{\tikzset{every picture/.style={line width=0.75pt}}
\begin{tikzpicture}[x=0.75pt, y=0.75pt, yscale=-1, xscale=1]
\draw (90,20) arc[start angle=0, end angle=180, radius=30] -- cycle;
\draw[line width=1.5] (20,20) -- (100,20);
\draw[dash pattern=on 4.5pt off 4.5pt] (70,20) -- (60,50);
\filldraw[fill=white, draw=black] (60,50) circle (3.67);
\end{tikzpicture}}}&= -i\lambda \kk_1\kk_2\int d\eta \frac{1}{\eta^2H^3}G_{-+}(k_1; \eta',\eta)G_{-+}(k_2; \eta',\eta)G_{-+}(q; \eta',\eta)
     \notag \\
    &= -\lambda \kk_1\kk_2\frac{H^3}{8k_1^3k_2^3q^3}e^{ik_T\eta'}\left[i\left(1-\eta'^2\sum_{i\neq j}k_ik_j\right)+\left(-k_T\eta'+k_1k_2q\eta'^3\right)\right]\times  \notag\\
    & \int \frac{e^{-ik_T\eta}}{\eta^2}\left[1+ik_T\eta-\eta^2\sum_{i\neq j}k_ik_j-i\eta^3k_1k_2k_3\right].
\end{align} 
Adopting the same conventions as before we evaluate the integral, and thanks to a cancellation (of the log-divergent term), we get   \begin{align}
    \text{Re}[int]= -e^{-ik_T\eta'}\left(\frac{1}{\eta'}+\frac{\eta'k_1k_2q}{k_T}\right) ,\notag \\
    \text{Im}[int]= - e^{-ik_T\eta'}\left(\frac{k_1k_2q}{k_T^2}+ \frac{\sum_{i\neq j}k_ik_j}{k_T}\right). 
\end{align}
To evaluate the bispectrum, we compute the now usual 
\begin{align}
    \braket{\sigma_{\kk_1}\sigma_{\kk_2}\chi_\qq}&= \lim_{\eta'\to 0}2\left[\text{Re}[int]\text{Re}[fac]-\text{Im}[int]\text{Im}[fac]\right]  \notag \\
    &= \frac{\lambda H^3}{4k_1^3k_2^3q^3}(\kk_1 \cdot \kk_2)\left(k_T-\left(\frac{k_1k_2q}{k_T^2}+ \frac{\sum_{i\neq j}k_ik_j}{k_T}\right)\right).
\end{align}
\paragraph{Fossils' limit}
As before, we now take the limit \begin{align}
    \lim_{\qq\to 0}\frac{\bar{\chi}\braket{\sigma_{\kk_1} \sigma_{\kk_2}\chi_\qq}}{\braket{\chi_\qq \chi_{-\qq}}}&=\bar{\chi}\frac{\lambda H^3}{4k_1^3k_2^3q^3}(\kk_1\cdot\kk_2)\left(\frac{3}{2}k_1\right) \frac{2q^3}{H^2} \notag \\
    &= \frac{H^2}{2k_1^3}\left(-\frac{3}{2}\frac{\lambda\bar{\chi}}{H}\right)
\end{align}
as $k_1, k_2\to k$, $k_T\to 2k_1$ and $\kk_1 \cdot \kk_2\to -k^2$ (we remark that the orientation of all the $\kk s $ in the triangle is such that the vectorial sum of momenta is null). We can appreciate that even in this case there is a precise correspondence with the non-perturbative result, cf. with Equation (\ref{np spaceder}).

\section{A concrete inflationary model beyond perturbation theory}
To show the versatility of this technique, we apply it in a concrete model of Inflation. We choose a model inspired by a single field slow-roll Inflation, parametrized by the slow-roll parameter $\epsilon$. We employ the same cubic Lagrangian of \cite{Maldacena_2003}, which is obtained from General Relativity with the additional scalar degree of freedom of the inflaton, and we work in the gauge in which the scalar curvature perturbation $\zeta$ and the tensor perturbation $\gamma_{ij}$ are explicit, as they are constant outside the horizon (cf. Eq. (3.1) in \cite{Maldacena_2003}). We set $M_{pl}=1$. We will focus on two scenarios: the one with mixed term $\gamma_{ij}\zeta\zeta$ - in which we are going to integrate out the long $\gamma_{ij}$ mode - and the other one with the $\gamma_{ij}\gamma_{ij}\zeta$ one - in which we are going to integrate out $\zeta$. We are going to expand to first order the non-perturbative results and check their equivalence with the Fossils' approach. The choice of this particular model is motivated by its simple nature, by the fact that many observables (e.g. the bispectra) are formally very well known, and by the fact that its couplings are common in other models of Inflation (e.g. \cite{Chen:2006nt}). We remain agnostic on the size of the amplitudes of the perturbations on scales different with respect to the ones we have access to, and therefore we treat $\bar{\gamma}^s, \bar{\zeta}$ as free parameters. 
\subsection{Integrating out one long tensor mode in \boldmath \texorpdfstring{$\gamma\zeta\zeta$}: interaction}
 The action reads: 
\begin{align}
    S= \int d\eta d^3\xx \bigg\{ \frac{\epsilon}{\eta^2H^2}\left[\zeta'\zeta'-(\partial_i\zeta)^2\right]+ \frac{1}{8\eta^2 H^2}\left[\gamma_{ij}'\gamma_{ij}'-(\partial_k \gamma_{ij})^2\right] \notag \\
    +\left[\epsilon \frac{1}{\eta^2H^2}\gamma_{ij}\partial_i\zeta \partial_j\zeta \right]\bigg\}.
\end{align}
A few remarks follow: 1) We are cherry-picking only the interacting term with two scalars and one graviton, and we are neglecting the other interacting terms that appear in the cubic action, such as the $\gamma \gamma \zeta$ (which will be considered in the next subsection) and $\gamma \gamma \gamma $ and $\zeta \zeta \zeta $ (which are not relevant for this approach as they involve just one field).This is justified by the fact that the other interacting terms contribute to the bispectrum $\braket{\zeta\zeta\gamma}$ only at loop level, and as pointed out in \cite{Celoria_2021} and as we will clarify in a dedicated section in this work, this non-perturbative approach resums only tree-level diagrams. 2) The $\zeta$ action (also the kinetic term) is of order $\epsilon$ because the curvature perturbation is well-defined only as long as we are not in pure de Sitter, ($\epsilon > 0$). 3) Dimensional analysis shows that $[\gamma_{ij}]=[\zeta]=0$ as they appear in the metric tensor, and the total mass-dimension in the action is kept by a $M_p^2$ factor that we omit. 4) We remind that the polarization tensor expansion reads: 
\begin{equation}
    \gamma_{ij}(\eta,\mathbf{x})=\int \frac{d^3\mathbf{k}}{(2\pi)^3}\sum_{s=+,\times}\epsilon^s_{ij}\gamma^s(\eta, \kk) e^{i\mathbf{k}\mathbf{x}}=\int \frac{d^3\mathbf{k}}{(2\pi)^3}\gamma_{ij}(\eta, \kk) e^{i\mathbf{k}\mathbf{x}}.
     \end{equation}
     Power spectra of free fields are defined using the shorthand notation $\gamma(\eta, \kk)_{ij}=\gamma_{\kk,ij}$, $\gamma^{s}(\eta, \kk)=\gamma_{\kk}^s$ and $\zeta(\eta, \kk)=\zeta_{\kk}$ in correlators, as before: 
    \begin{equation}
    \braket{\zeta_{\kk_1}\zeta_{\kk_2}}=(2\pi)^3\delta^{(3)}(\mathbf{k}_1+\mathbf{k}_2)\frac{H^2}{4\epsilon k_1^3},
\end{equation}
\begin{equation}
    \braket{\gamma_{\kk_1}^s\gamma_{\kk_2}^{s'}}=(2\pi)^3\delta^{(3)}(\mathbf{k}_1+\mathbf{k}_2)\delta_{ss'}\frac{H^2}{k_1^3},
\end{equation}
\begin{equation}
  \braket{\gamma_{\kk_1, ij}\gamma_{\kk_2,ij}}=(2\pi)^3\delta^{(3)}(\mathbf{k}_1+\mathbf{k}_2')\epsilon_{ij}^s\epsilon_{ij}^{s'}\braket{\gamma_{\kk_1}^s\gamma_{\kk_2}^{s'}}=(2\pi)^3\delta^{(3)}(\mathbf{k}_1+\mathbf{k}_2)\frac{4H^2}{k_1^3},
\end{equation}
given that $\epsilon_{ij}^s\epsilon_{ij}^{s'}=2\delta_{ss'}$. A final remark is regarding notation. We will refer to $\kk_j$ as the $j$ component of the three dimensional vector $\kk$, not to be confused with $\kk_1, \kk_2$ which are three dimensional vectors, or with $k_1, k_2$ which are the moduli of such vectors. Whenever two three dimensional vector are close to each other we intend a scalar product, indicated as $\cdot$, as in the previous sections, while for vector-tensor contractions we adopt indices notation. We can now start with the evaluation of the two equations of motion, the one for $\gamma_{ij}$ which reads: \begin{equation}
    \gamma_{ij}''-\frac{2}{\eta}\gamma_{ij}'+k^2\gamma_{ij}=-\frac{\epsilon}{H}\int \frac{d^3\kk}{(2\pi)^3}\kk\cdot(\qq-\kk)\zeta(\kk)\zeta(\qq-\kk),
\end{equation}
and the equation for $\zeta$ which reads: \begin{align}
    \zeta''-\frac{2}{\eta}\zeta'+k^2\zeta + \frac{1}{H}\int \frac{d^3\qq}{(2\pi)^3}(\qq-\kk)_i(\qq-\kk)_j\zeta(\kk-\qq)\gamma_{ij}(\qq) \notag \\
    -\frac{1}{H}\int \frac{d^3\qq}{(2\pi)^3}\qq_i(\kk-\qq)_j\zeta(\kk-\qq)\gamma_{ij}(\qq)=0\, .
\end{align}
This case is very close to the one with spatial derivatives, treated in the previous section. We can simplify the system assuming that the slow-roll parameter $\epsilon$ makes the equation for $\gamma$ almost free (since the term on the R.H.S. is slow-roll suppressed with respect to all the others and hence at the end it would lead to a sub-leading contribution). The solution is a free tensor in de Sitter space: 
\begin{equation}
    \gamma_{ij}(\kk)=\delta^{(3)}(\kk-\kk_\gamma)(2\pi)^3\bar{\gamma}_{ij}(1-ik\eta)e^{ik\eta}.
\end{equation}
Once we plug this equation into the one for $\zeta$, with the additional hypothesis that the hierarchy in momenta is $k_\gamma \ll k_\zeta$, we can get in the late time limit: \begin{equation}
    \zeta''-\frac{2}{\eta}\zeta'+(k^2+\kk_i\kk_j\bar{\gamma}_{ij})\zeta=0 .
\end{equation}
This equation shows a modification of the kinetic term, and comes from an effective action \begin{equation}
    S_\text{eff}=\int d\eta d^3\xx  \frac{\epsilon}{\eta^2H^2}\left[\zeta'\zeta' -\partial_i\zeta\partial_j\zeta (\delta_{ij}-\bar{\gamma}_{ij})\right].
\end{equation} 
We notice as expected that the action has a non-canonical kinetic term. The modification is not a mere reduction of the speed of sound as was in the previous case, but a warp of the spatial metric due to the large tensor mode. After switching to momentum space \begin{equation}
     S_\text{eff}=\int d\eta \frac{d^3\kk}{(2\pi)^3}  \frac{\epsilon}{\eta^2H^2}\left[\zeta'\zeta' -\zeta\zeta (\delta_{ij}-\bar{\gamma}_{ij})\kk_i\kk_j\right],
\end{equation} and rescaling the field \begin{equation}
    Z=\zeta\sqrt{\frac{(\delta_{ij}-\bar{\gamma}_{ij})\kk_i\kk_j}{k^2}} ,
\end{equation} we can extract the power spectrum, which reads \begin{equation}\label{np zeta ps}
    \braket{\zeta_{\kk_1} \zeta_{\kk_2}}_\text{eff}=\braket{\zeta_{\kk_1}\zeta_{\kk_2}}\left(\frac{k_1^2}{k_1^2-\bar{\gamma}_{ij}\kk_i \kk_j}\right)^{3/2}.
    \end{equation}
    This quantity, which has been obtained beyond perturbation theory, can be expanded to first order in $\epsilon$: 
    \begin{equation}
     \braket{\zeta_{\kk_1} \zeta_{\kk_2}}_\text{eff}\sim \braket{\zeta_{\kk_1}\zeta_{\kk_2}}\left(1+\frac{3}{2}\bar{\gamma}_{ij}\frac{\kk_i\kk_j}{{k_1^2}}\right). 
\end{equation}
\paragraph{Bispectrum}
The bispectrum of this theory is a well-known result in literature, first computed in \cite{Maldacena_2003}, and we can skip the computation. Applying the Fossils' prescription we have:
\begin{align}
\braket{\zeta_{\kk_1}\zeta_{\kk_2}}|_{\gamma_L}&=\braket{\zeta_{\kk_1}\zeta_{\kk_2}}+\frac{\bar{\gamma}^s\braket{\zeta_{\kk_1}\zeta_{\kk_2}\gamma^s_\qq}}{\braket{\gamma^s_\qq\gamma^s_{-\qq}}'}=  \notag \\
&= \braket{\zeta_{\kk_1}\zeta_{\kk_2}} + \bar{\gamma}^s \frac{q^3\epsilon_{ij}^s}{H^2}\frac{\kk_i\kk_j}{k_1^2}\frac{3H^4}{8\epsilon k_1^3q^3}(2\pi)^3\delta^{(3)}(\mathbf{k}_1+\mathbf{k}_2)  \notag \\
&= \braket{\zeta_{\kk_1}\zeta_{\kk_2}}\left(1+\frac{3}{2}\frac{\bar{\gamma}_{ij}\kk_i\kk_j}{k_1^2}\right).
\end{align}
We see a perfect match at first order with Eq. (\ref{np zeta ps}), meaning that the procedure beyond perturbation theory correctly describes the perturbative result. We can see how having accounted for the effect of the long tensor mode with the Fossils' approach introduces a quadrupole (proportional to $\hat{\kk}_i\hat{\kk}_j$) distortion in the scalar power spectrum \cite{Jeong:2012df, Dimastrogiovanni:2012st, Dimastrogiovanni:2014ina, Dimastrogiovanni:2022afr}.
\subsection{Integrating out one long scalar mode in \boldmath \texorpdfstring{$\gamma\gamma\zeta$}: interaction}
In this subsection we will postpone the lengthier calculations to Appendix C, but the procedure is extremely similar to the ones we just carried out. We will summarize here the main steps. 
\newline 
The starting action reads, keeping explicit the dependence from the metric tensor:\footnote{In this case as well the same logic applies: all the other contributions coming from cubic terms in the action such as $\gamma \gamma \zeta$, $\gamma \gamma \gamma $ and $\zeta \zeta \zeta $ to the bispectrum of $\braket{\zeta \gamma_{ij}\gamma_{ij}}$  arise only at one loop order.} \begin{align}
    S=\int d^3\xx d\eta \bigg\{\frac{-\epsilon}{\eta^4H^4}\left[\zeta'\zeta'g^{00}+\partial_i\zeta \partial_j\zeta g^{ij}\right]-\frac{1}{8\eta^4H^4}\left[\gamma_{ij}'\gamma_{ij}'g^{00}+\partial_k\gamma_{ij}\partial_h\gamma_{ij}g^{hk}\right] \notag \\
    \frac{-\epsilon}{8\eta^4H^4}\left[\zeta \gamma_{ij}'\gamma_{ij}'g^{00}-\zeta\partial_k\gamma_{ij}\partial_h\gamma_{ij}g^{hk}\right]-\frac{1}{4\eta^4H^4}g^{hk}\left[\gamma_{ij}'\partial_k\gamma_{ij}{\partial_h}\chi\right],
\end{align}
where the first line is the usual quadratic action, and the second line is the interacting action. We can notice the term that depends on the auxiliary field $\chi$\footnote{We refer the reader to \cite{Maldacena_2003} for the derivation of this Lagrangian.}, that on shell respects the equations (in real and momentum space): \begin{equation}
    \nabla^2 \chi = \epsilon \zeta', \; \; q^2\chi = -\epsilon \zeta',
\end{equation} and that therefore is of the same order in slow-roll parameters as the other interacting term. \newline 
After computing the equations of motion for $\gamma$ and $\zeta$, and the constraint equation for the auxiliary field $\chi$), there exists a regime in which: 1) $\gamma \ll \zeta$, so that the $\zeta$ equation is an equation for a free and massless scalar in de Sitter, and we can insert its solution into the $\gamma$ equation; 2) $\kk_\zeta \ll\kk_\gamma$ so that the late time limit exists and an effective action which is a functional of the sole $\gamma$ can be derived in such a late time limit; 3) $\bar{\zeta}\epsilon \sim 1$ so that contributions from perturbation theory that are $\epsilon$ suppressed but are enhanced by external $\zeta$ legs are still felt by $\gamma$ in the bulk. \newline  The resulting equation of motion for $\gamma$ reads  \begin{equation}\label{eom zeta}
    \left(\gamma''_{ij}-\frac{2}{\eta}\gamma_{ij}'\right)(1+\epsilon \bar{\zeta})+k^2\gamma_{ij}(1-\epsilon \bar{\zeta})=0\, . 
\end{equation}
 The corresponding effective action reads: \begin{equation}
    S_{\text{eff}}=\int d\eta d^3\xx \frac{-1}{8\eta^4H^4}\left[\gamma'_{ij}\gamma'_{ij}g^{00}(1+\epsilon\bar{\zeta})+\partial_h\gamma_{ij}\partial_k\gamma_{ij}g^{hk}(1-\epsilon\bar{\zeta})\right],
\end{equation}where we see that\begin{equation}
    c_s=\sqrt{\frac{1-\epsilon\bar{\zeta}}{1+\epsilon\bar{\zeta}}}.
\end{equation} If we normalize $\gamma_{ij}$ to extract swiftly the power spectrum: 

\begin{equation}
    G_{ij}=\gamma_{ij}\sqrt{1-\epsilon\bar{\zeta}},
\end{equation} we get 
\begin{equation}
    \braket{G_{\kk_1, ij} G_{\kk_2, ij}}_\text{eff}'= \frac{4H^2}{k_1^3}\sqrt{\frac{1+\epsilon\bar{\zeta}}{1-\epsilon\bar{\zeta}}}=(1-\epsilon\bar{\zeta})\braket{\gamma_{\kk_1, ij}\gamma_{\kk_2, ij}}_\text{eff}',
\end{equation}and therefore, moving to polarization ($s,s'$) basis:   
\begin{equation}\label{np gamma ps}
    \braket{\gamma_{\kk_1}^s\gamma_{\kk_2}^{s'}}_\text{eff}'=\delta_{ss'}\frac{H^2}{k_1^3}\frac{\sqrt{1+\epsilon\bar{\zeta}}}{(1-\epsilon\bar{\zeta})^{3/2}}. 
\end{equation}
As before, this expression, valid beyond perturbation theory, can be expanded at first order to obtain \begin{equation}
    \braket{\gamma_{\kk_1}^s\gamma_{\kk_2}^{s'}}_\text{eff}'\sim \delta_{ss'}\frac{H^2}{k_1^3}\left(1+2\epsilon\bar{\zeta}\right)= \braket{\gamma_{\kk_1}^s\gamma_{\kk_2}^{s'}}'\left(1+2\epsilon\bar{\zeta}\right). 
\end{equation}
\paragraph{Bispectrum}
Again we can write the bispectrum, which has been computed in \cite{Maldacena_2003}, and insert it in the Fossils' prescription. The Fossils' prescription reads \begin{align}
\braket{\gamma_{\kk_1}^s\gamma_{\kk_2}^{s'}}|_{\zeta_L}&=\braket{\gamma_{\kk_1}^s\gamma^{s'}_{\kk_2}}+(2\pi)^3\delta^{(3)}(\mathbf{k}_1+\mathbf{k}_2')\frac{\bar{\zeta}\braket{\zeta_\qq\gamma^s_{\kk_1}\gamma^{s'}_{\kk_2}}}{\braket{\zeta_\qq\zeta_{-\qq}}'} \notag \\
&= \braket{\gamma_{\kk_1}^s\gamma^{s'}_{\kk_2}}\left(1+2\epsilon\bar{\zeta}\right).
\end{align}

\section{A model violating the consistency conditions}
A famous theorem by Weinberg on single field inflationary models \cite{Weinberg_2003} states that all cosmological
models in FLRW spacetime have two adiabatic modes of fluctuation, one of which has
constant and identical curvature perturbation on comoving and on uniform-density hypersurfaces ($\zeta$ and $\mathcal{R}$) on large scales, while the other has $\zeta = \mathcal{R}= 0$. Closely related to this theorem are the \vg consistency conditions'' \cite{Maldacena_2003}, namely that the squeezed limit of the bispectrum of all single field models of Inflation should be obtained by a combination of the power spectrum and of its tilt, schematically: \begin{equation}
    \lim_{\kk_1 \to 0}\braket{\zeta_{\kk_1} \zeta_{\kk_2} \zeta_{\kk_3}}= -(n_s-1) \braket{\zeta_{\kk_2} \zeta_{\kk_3}}\braket{\zeta_{\kk_2} \zeta_{\kk_3}},
\end{equation} where the scalar tilt is defined through $\braket{\zeta_{\kk_1} \zeta_{\kk_2}}\sim k_1^{-3+n_s}$. Since the formulation of such conditions, many attempts at finding ways to evade them have been explored. These explorations were motivated by the theoretical pursuit of deepening the understanding of the underlying assumptions of such consistency conditions, and by the phenomenological observation that the possibility of observing a bispectrum in the squeezed limit, if it is ruled by a very small $(n_s-1)=2\eta -6\epsilon$ parameters, would be out of reach for current cosmological surveys. The nature of the consistency condition is linked to the fact that, in a bispectrum, a very long mode of the scalar perturbation $\zeta$, which is frozen by the time the other two cross the horizon, can be traded for a small time dilation an thus be interpreted as slightly retarding or anticipating the horizon exit of the remaining two modes. This is represented in the squeezed bispectrum by the tilt of the scalar power spectrum, encoding the breaking of scale invariance. Some noticeable ways out from the consistency condition have indeed been found, among which are the so-called Solid Inflation models (\cite{Endlich_2013} and related works) and the Supersolid Inflation models (\cite{Celoria_2017, Celoria_2021_denis}), where the consistency condition is evaded because the symmetry breaking pattern of the theory does not admit adiabatic modes, or some more generic inflationary models that do admit non-attractor solutions such as \cite{Chen_2013}, where $\zeta$ is not conserved super horizon, leading to a breaking of the consistency conditions\footnote{For a treatment and comparison of such bispectra in the Fossils' approach see \cite{Dimastrogiovanni:2022afr}.} (see \cite{Chen:2010xka} for a a detailed discussion about various possibilities to violate the consistency relations).  
While this work, despite dealing with long modes in inflation, is not strictly linked to the consistency conditions, one might wonder if it remains applicable also in such a scenario. This is in fact the case: if the theory displays a breaking of the consistency conditions perturbatively at tree level (that is, in the Fossils' procedure), this breaking will be present also in the procedure beyond perturbation theory. We provide an example. \subsection{Beyond perturbation theory with a large bispectrum}
Following the nice work in \cite{Chen_2013}, we are provided with a Lagrangian for the curvature perturbation $\zeta$ (dubbed $\mathcal{R}$ in the original text) that predicts a large bispectrum in the squeezed (local) configuration, namely \begin{equation}
    f_\text{NL}^\text{loc}\sim \frac{5}{4c_s^2},
\end{equation}
so the $f_\text{NL}^\text{loc}$ is enhanced by one over the speed of sound squared. The quadratic action for the general single field inflation with non-canonical kinetic term to quadratic order reads: \begin{align}
    S_2&=\int d \eta d^3 \xx\left\{-\frac{1}{2} \sqrt{-g}\left[\left(\frac{\epsilon}{c_s^2}\right)\zeta' \zeta'  +\epsilon \partial_i\zeta \partial_i \zeta\right]\right\}.
\end{align} The cubic one, in the variable $\zeta_n$, which is a field redefinition of $\zeta$ defined as\footnote{Actually, the field redefinition contains many more terms which we have omitted, as they involve at least one derivative of $\zeta$, and they should generate terms which are vanishing superhorizon \cite{Arroja_2011}. For a recent treatment of the field redefinitions, which compares different approaches and proposes to put them in the context of canonical transformation in the Hamiltonian formalism, see \cite{Braglia_2024}, and for previous works see e.g. \cite{Chen_2007, Arroja_2011}. We have checked that our results agree with the former approach.} \begin{equation}\label{field redef}
    \zeta = \zeta_n+\frac{\eta}{4c_s^2}\zeta_n^2+\dots ,
\end{equation}
reads: \begin{align}
    S_3=\int d \eta d^3 \xx \bigg[-\frac{1}{2}\sqrt{-g}\bigg[  \left(\frac{\epsilon}{c_s^2}\right)(c_s^2-1)\zeta_n \partial_i\zeta_n \partial_i\zeta_n - \left(\frac{\epsilon}{c_s^4}(3-3c_s^2) \right)\zeta_n \zeta_n' \zeta_n' \nonumber \\ - (\Sigma\left(1-\frac{1}{c_s^2}\right) +2\lambda) \frac{\zeta_n^{'3}}{H^3}  \bigg], 
\end{align}
where for brevity we dubbed $\zeta_n' = \partial_0\zeta_n$ and where repeated indices are summed over omitting the $g^{\mu \nu}$. In the above equations $\epsilon , \eta$ are the first two slow-roll parameters (the latter not to be confused with conformal time), and $c_s, \Sigma, \lambda$ are other parameters of the theory, which can be taken as constants. We can read that the kinetic term displays a modification of the speed of sound. From the above $S_3$, and adding the contributions to the bispectrum of the field redefinition, it is possible to compute the value of $f_\text{NL}^\text{loc}$. \newline As in the previous cases, our strategy will be to compute the effective action for the short modes of $\zeta_n$, in presence of a large and long mode. In order to obtain it consistently, we must consider the effect that the field redefinition has in modifying the quadratic action as well, without integrating by parts and using the linear equations of motion to simplify the action, as this trick somewhat requires a perturbative distinction between linear and non linear EoM, which would spoil any later attempt to go beyond perturbation theory. The resulting action reads: \begin{align}
    S&=\int d \eta d^3 \xx \bigg[-\frac{1}{2}\sqrt{-g}\bigg(\left(\frac{\epsilon}{c_s^2}\right)\zeta_n' \zeta_n'  +\epsilon \partial_i\zeta_n \partial_i \zeta_n  \nonumber \\ 
    &-  \left(\frac{\epsilon}{c_s^2}\right)(1-c_s^2)\zeta_n \partial_i\zeta_n \partial_i\zeta_n + \left(\frac{\epsilon}{c_s^4}(-3+3c_s^2) \right)\zeta_n \zeta_n' \zeta_n' - (\Sigma\left(1-\frac{1}{c_s^2}\right) +2\lambda) \frac{\zeta_n^{'3}}{H^3}    \nonumber \\ 
    & +\frac{\epsilon}{c_s^2} \left(\frac{\eta}{c_s^2}\zeta_n'^2\zeta_n \right) + \epsilon\left(\frac{\eta}{c_s^2}\zeta_n(\partial_i \zeta_n)^2 \right) \bigg)\bigg], 
\end{align} where in the third line it is evident the effect of the field redefinition. If we imagine a very large and long wavelength perturbation\footnote{We remark that we are treating a large and long mode of $\zeta_n$ on the same footing as a large and long mode of $\zeta$, as they coincide at first order, and we are interested in obtaining the bispectrum, which contains only one long mode.} of the $\zeta_n$ field, dubbed $\zeta_L$, derive the equations of motion for it, and and insert them in the action, the effective action resulting for the leftover short perturbations is: 
\begin{align}
    S_\text{eff}  =\int d \eta d^3 \xx & \bigg[- \frac{1}{2}\sqrt{-g}\bigg( \frac{\epsilon}{c_s^2}\zeta_n' \zeta_n' \left(1 + \zeta_L\left( \frac{-3+3 c_s^2}{c_s^2}+ \frac{\eta}{c_s^2} \right)\right)   \nonumber \\ & +\epsilon \partial_i\zeta_n \partial_i \zeta_n \left( 1+ \zeta_L\left(\frac{c_s^2-1}{c_s^2} + \frac{\eta}{c_s^2} \right)\right)    \bigg)\bigg], 
\end{align} as every other term is suppressed by derivatives of the long mode. If we renormalize the field $\zeta$ in order to isolate the effective speed of sound, using\begin{equation}
    Q = \zeta_n\sqrt{  1+ \zeta_L\left(\frac{c_s^2-1}{c_s^2} + \frac{\eta}{c_s^2} \right)  }, 
\end{equation} the action for the field $Q$ becomes
\begin{align}
      S_\text{eff}  =& \int d \eta d^3 \xx  \bigg[-\frac{1}{2}\sqrt{-g}  \\ & \bigg( \frac{\epsilon}{c_s^2}Q' Q' \left(1 + \zeta_L\left( \frac{-3+3 c_s^2}{c_s^2}+ \frac{\eta}{c_s^2} \right)\right)\left( 1+ \zeta_L\left(\frac{c_s^2-1}{c_s^2} + \frac{\eta}{c_s^2} \right)\right)^{-1}  +\epsilon \partial_i Q \partial_i Q    \bigg)\bigg], \nonumber
\end{align} or equivalently \begin{equation}
      S_\text{eff}  =\int d \eta d^3 \xx \bigg[\sqrt{-g} \frac{-\epsilon}{2}\bigg( \frac{1}{\hat{c}_s^2}Q' Q'    +\partial_i Q \partial_i Q    \bigg)\bigg], 
\end{equation}
in which we have isolated the effective speed of sound for the field $Q$:\begin{equation}
    \frac{1}{\hat{c}_s^2} = \frac{1}{c_s^2}\left(1 + \zeta_L\left( \frac{-3+3 c_s^2}{c_s^2}+ \frac{\eta}{c_s^2} \right)\right)\left( 1+ \zeta_L\left(\frac{c_s^2-1}{c_s^2} + \frac{\eta}{c_s^2} \right)\right)^{-1} .
\end{equation}
As usual, the power spectrum for the field $Q$ reads \begin{equation}
    \braket{Q_{\kk_1}Q_{\kk_2}}_\text{eff} = \frac{H^2}{2k^3\epsilon \hat{c}_s}. 
\end{equation} Combining this result with the definition of $\hat{c}_s$ and with the definition of the field $Q$, one can obtain the effective power spectrum in $\zeta_n$. The latter, once expanded to first order in the long mode reads: \begin{equation}
    \braket{\zeta_{n, \kk_1}\zeta_{n, \kk_2}}_\text{eff} = \frac{H^2}{2k^3\epsilon {c}_s}\left(1-\frac{\eta \zeta_L}{c_s^2}+...\right). 
\end{equation}At this point we can resort to perturbation theory to recover the bispectrum for the variable $\zeta$. We must exploit that, symbolically for our field redefinition: \begin{equation}
    \braket{\zeta_{\kk_1} \zeta_{\kk_2} \zeta_{\kk_3}} = \braket{\zeta_{n, \kk_1} \zeta_{n, \kk_2} \zeta_{n, \kk_3}} + \frac{\eta}{4c_s^2}[\braket{\zeta_{n, \kk_1}\zeta_{n, \kk_2}}\braket{\zeta_{n, \kk_1}\zeta_{n, \kk_3}} + \text{perm}],
\end{equation}
so in order to get the desired $ \braket{\zeta_{\kk_1} \zeta_{\kk_2} \zeta_{\kk_3}}$ we must first obtain $\braket{\zeta_n\zeta_n\zeta_n}$. The latter is obtained from the expansion (being more schematic here to ease the notation):
\begin{equation}
    \braket{\zeta_n\zeta_n}_\text{eff} = \braket{\zeta_n\zeta_n}(1-\eta\zeta_L/c_s^2+...) = \braket{\zeta_n\zeta_n}\left( 1+ \frac{\zeta_L \braket{\zeta_n\zeta_n\zeta_n}}{\braket{\zeta_n\zeta_n}^2} + ...\right), 
\end{equation}
where the power spectra of $\zeta$ and $\zeta_n$ are equal, as one cannot go at more than order 1 in the field redefinition. Plugging in $\eta = -6 $ as done in \cite{Chen_2013}, and accounting for the three permutations we get, to first order,\begin{equation}
    \braket{\zeta \zeta \zeta} = \braket{\zeta \zeta}^2\frac{3}{2c_s^2}. 
\end{equation} Accounting for the standard definition, omitting the delta functions, \begin{equation}
    \braket{\zeta_{\kk_1} \zeta_{\kk_2} \zeta_{\kk_3}}' = \frac{6}{5}f_\text{NL}( \braket{\zeta_{\kk_1} \zeta_{\kk_2}}^2 +\text{perm}), 
\end{equation} where we remark that this time there is only one configuration non vanishing, as we are in the squeezed limit to get the local configuration, so the permutation is just one, gives \begin{equation}
    f_\text{NL}^\text{loc}= \frac{5}{4c_s^2}, 
\end{equation} which is the factor computed in \cite{Chen_2013} and shows an amplification of the local bispectrum, evading the consistency conditions.
\section{Resumming diagrams}
\begin{figure}
    \centering
    \includegraphics{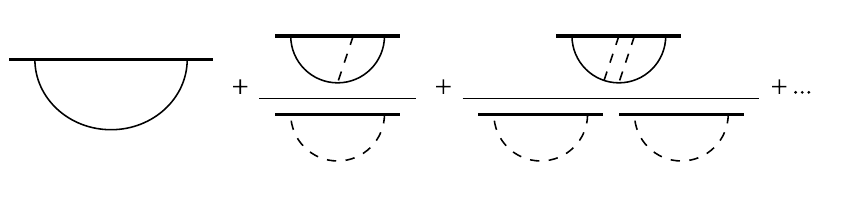}
    \caption{A diagrammatical sketch of what we mean by resummation. The leftmost part is the free power spectrum of the $\sigma$-like field, in the middle we display the Fossils' correction, namely the bispectrum normalized by the power spectrum of the $\chi$-like field, and in the rightmost part the first of the infinitely many corrections that are not considered in the Fossils' approach, but that this non-perturbative technique does resum. We clarify that the thin horizontal lines are simply fraction lines. }
    \label{fig: series}
\end{figure}

In this section, expanding on the work in \cite{Celoria_2021}, we clarify what class of diagrams we are resumming in the non-perturbative approach. The models we explored in this work are theories in which fields couple (modulo derivatives) like $\sigma^2\chi$ in the action, and we worked with the aim of refining the two point function of the $\sigma$-like field, in presence of a long and large mode $\bar{\chi}$. For the sake of brevity, in the following we will refer sometimes to the \vg $\sigma$-like'' and the \vg $\chi$-like'' field as the one quadratic and linear in the interaction term, respectively. \begin{figure}
    \centering
    \includegraphics[width= 1\linewidth]{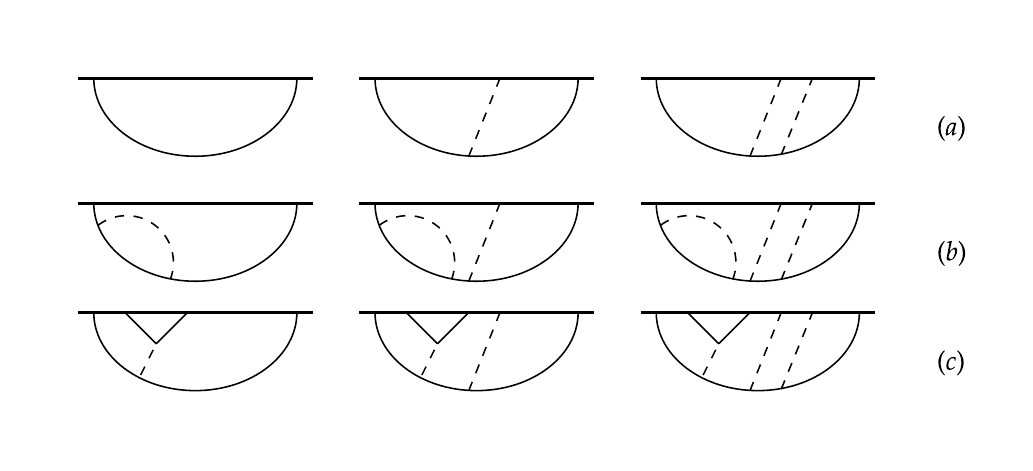}
    \caption{The thick horizontal line stands for the spatial hypersurface at the end of Inflation, and in the bulk full legs represent the $\sigma$-like field, dashed legs the $\chi$-like field. (a) Tree-level diagrams that are resummed in the effective theory thanks to the enhancement by the dashed legs (b) one-loop diagrams that are not resummed in the effective theory due to the suppression given by the mismatch between couplings and dashed legs, (c) tree-level diagrams that have more than two full legs, no matter the amount of dashed legs, are not resummed in the effective theory due to the excess of coupling with respect to the number of dashed legs. The effective theory is a free (i.e. with quadratic lagrangian) classical (i.e. without loops) theory for the $\sigma$-like field, and does not contemplate higher order correlators.}
    \label{fig: sum}
\end{figure}
\paragraph{Toy models}
In the toy models studied in this work, the small coupling $\lambda$ ruled the perturbative expansion. We however have been able to consider a scenario in which a large amplitude and long wavelength mode of $\chi$ was able to counterbalance the small coupling expansion. In this scenario we derived (without ever expanding in perturbation theory) a quadratic effective Lagrangian in the leftover $\sigma$ field, once the $\chi$ long-mode has been integrated-out. This lagrangian is quadratic, and therefore only predicts a nontrivial two point function. This two point function is somewhat sensitive to a class of infinitely many contributions from perturbation theory, displayed in Figure \ref{fig: sum}(a), where we can imagine to navigate the series in which for every small coupling, a large $\bar{\chi}$ counterbalances the naive order of magnitude of the contribution. We must however be careful, as the effective theory is not sensitive to the \vg whole'' bunch of two $\sigma$ lines diagrams, since all loop diagrams are (on a naive dimensional analysis line of thought, hence neglecting possible particular effects or enhancements that might arise in at loop level in some peculiar theory) further suppressed by higher powers of the small coupling, and for this reason are not resummed in the effective action. Looking at the first two diagrams from the left in Figure \ref{fig: series} we can see the Fossils' correction: to the power spectrum (the leftmost diagram) it is added the squeezed limit of the bispectrum, normalized by the power spectrum of the long $\chi$ mode (the diagrams at the center). The Fossils' approach stops at first order, hence it does not consider the rightmost diagram. We remark that the non-perturbative technique considered in this work, despite working \vg only'' at tree level, is not merely resumming algebraic contributions. In fact, in the in-in formalism - due to the nature of de Sitter space, or of any spacetime not enjoying the four Lorentz translational symmetries - every interacting vertex represents a time integral in the final correlator, hence every step in the ladder of perturbation theory is a nested integral to compute. In this sense, this non-perturbative technique is a remarkable simplification. 
\paragraph{Concrete model}
We also briefly comment on what happens in the model inspired by single field slow-roll Inflation we analyzed. We have to keep in mind that, differently from before where all power spectra were of order 0 in the coupling, in this model $\braket{\zeta \zeta}\sim \epsilon^{-1}$. Hence, looking again at Figure \ref{fig: series}, where now the dashed line represents the tensor mode, every vertex is still an $\epsilon$ contribution, but also every full line is a $\epsilon^{-1}$ contribution, making the whole series of diagrams of order $\epsilon^{-1}$, and making the $\gamma_{ij}$ insertion stand alone, without $\epsilon$ insertions, in the Taylor expansion. The same reasoning applies for the other two lines of Figure \ref{fig: sum}: adding loops amounts to increasing the number of $\epsilon$ insertions which are not counterbalanced by internal $\gamma$ legs, and considering higher order correlators in $\zeta$ is meaningless as the effective theory is intrinsically quadratic in $\zeta$. \newline 
For the $\gamma \gamma \zeta$ term, we can again look at Figure \ref{fig: series}, this time dashed lines are $\zeta$ fields. The leftmost diagram is of order $\epsilon^0$, the Fossils' correction is naively of order $\epsilon^1$ ($\epsilon$ coming from the vertex, $1/\epsilon$ from the $\zeta$ propagator in the bispectrum and $\epsilon$ from the the $\zeta$ propagator in the power spectrum), but is enhanced by the asymptotic value $\bar{\zeta}\sim 1/\epsilon$. One can see that the same rule applies also for the second fraction, of order $\epsilon^2\bar{\zeta}^2$: the whole series is of order $\epsilon^0$ and is therefore resummed in the non-perturbative approach. 
\subsection{Comparison with existing literature}
A complete and comprehensive comparison of the technique introduced in \cite{Celoria_2021} and on which this work builds upon with all existing literature is certainly beyond the scope of this paragraph. It is however instructive to point out the differences between this approach and other approaches that work beyond perturbation theory, or that employ techniques of Effective Field Theory (EFT) to integrate out fields. To conclude, we will draw some consideration to broadly frame the techniques treated in this work with some of the literature of the so-called \vg separate Universe'' approach. \newline 
Beginning with the former techniques, the main difference of the approach carried out here with respect to other recent resummation techniques that reach beyond perturbation theory at tree level in the inflationary era, such as the one in Appendix A.2 of \cite{Bordin_2018},  or the works of \cite{Werth:2023pfl, Pinol:2023oux} is the nature of the interaction: such works consider quadratic mixing between two fields, and resum many oscillations in an effective propagator, whereas in \cite{Celoria_2021} and here a cubic vertex is considered, and resummed in the effective propagator. Other recent contributions, such as \cite{Palma_2025, Huenupi_2024} and related works) focus on resumming IR loop effects in de Sitter, which in the specific diagrammatical configurations considered should give logarithmic corrections, and for this reason are different and may be complementary to the approach carried out here. It would be interesting to see if this approach can be combined with any of the above mentioned ones, and to understand to which extent the results these approaches give in terms of modifications of the power spectra are qualitatively equivalent, or if instead they have distinct signatures.
\newline   
Regarding the second techniques, a remarkable effort has been done in the direction of deriving effective behavior of light fields (particularly, the inflaton), in cases in which a heavy field is integrated out in an EFT fashion (\cite{Rubin_2001, Tolley:2009fg, Dimastrogiovanni:2012st, Ach_carro_2012, Ach_carro_2015} and related works). The technique considered in this paper can share the final result with some of the above mentioned ones (e.g. it is possible to introduce an effectively lower speed of sound), but differs in one crucial part: it needs not \textit{heavy} fields to be integrated out, rather \textit{light} - ideally massless - fields to be integrated out. We stress in fact that the effective action is not a \textit{low energy} effective one, but a \textit{late time} effective one, and to be viable it needs a non-dynamical field, which solution has to be plugged in the action, a condition that is indeed satisfied asymptotically (and non trivially) at late times by massless fields, indeed in the spirit of the Fossils' approach. \newline 
We conclude this section by briefly commenting on possible connections with  the so-called \vg separate Universe'' approach \cite{Wands_2000,Wands_2010,Byrnes_2006} and related works, and drawing a comparison with the techniques treated here. Such an approach, which despite dealing with long modes in Inflation is quite different from the one treated here, focuses on the curvature perturbation and its superhorizon evolution. It hinges on a separation of scales $\lambda_0 \gg \lambda \gg \lambda_s$. The background is assumed to be at very large scales $\lambda_0$, while on smaller scales, in patches of size $\lambda_s$ (still of the order or larger than the Hubble horizon) the Universe can be treated as homogeneous, but its state may vary from patch to patch independently, as if they were separate Universes. Linking together the profile of the curvature perturbation on different patches leads to its determination on the scale $\lambda$, in which it may be oscillating with respect to the background, and which fluctuations are ultimately linked with the variations in the number of e-folds $\delta N$. We remark that this latter quantity in principle is not an expansion in small couplings of the theory, and for this reason it is claimed that this formalism can reach beyond perturbation theory. Using such an approach, it has been remarkably proven that the curvature perturbation at tree level is conserved on superhorizon scales, in presence of sole adiabatic perturbations \cite{Wands_2000} (see also~\cite{Lyth_2005} but for previous works, see~\cite{Salopek:1990jq}). We now come to the comparison between the two methods, focusing on the hypothesis they are built upon. The \vg separate Universe'', in order to compute correlation functions, exploits the so-called gradient expansion, which retains only large scale modes \cite{Lyth_2005}, and which (only on large scales) is equivalent to (quantized) cosmological perturbation theory \cite{Lyth_2005_ng}, the latter being valid at all times. In turn cosmological perturbation theory has been proven equivalent to the in-in formalism, as shown in \cite{Seery_2008}. Given that the technique presented in this work, once expanded order by order in perturbation theory, is equivalent to the in-in formalism, one might argue that it should give the same results of the  \vg separate Universe'' (on superhorizon scales) if in the latter approach one collects all the corresponding powers of the coupling. On general grounds, however, the resummations, or the results beyond perturbation theory that the two approaches provide, if not compared order by order in the coupling, seem not granted to be equivalent, being resummations of a different kind of expansion. In order to establish a complete equivalence between the models beyond any doubt one should perform the same computation in two scenarios, which we believe goes beyond the scope of this work, and we leave for future explorations. To conclude we also stress that the techniques treated in this work are not focusing on the curvature perturbation alone, but are completely general, as it was shown in the examples regarding the toy models.

\section{Conclusions}
The main result of this work is the formalization of the missing link between the approach of the inflationary Fossils' and the technique introduced in \cite{Celoria_2021}, capable of reaching beyond perturbation theory. This result can be understood both as a generalization of the Fossils' approach and as a first order check of the validity of the result beyond perturbation theory. To conclude, we list some remarks and possible future directions. 
\begin{itemize}
    \item The Fossils' approach is fully justified with, generally speaking, different assumptions than the ones we make (it suffices to have a very \textit{long} mode of the Fossils' field). It remains therefore a perfectly viable approach to investigate relic fields in the early Universe, and in this sense we are not \vg correcting'' anything wrong. The non-perturbative technique we are considering in this paper, under the assumptions that make it applicable, is capable of extending the Fossils' prescription to include contributions from all orders in perturbation theory. These contributions are not increasingly small, due to the enhancement given by the fact that the Fossils' mode can be also \textit{much larger} than its ``normal'' fluctuation. Moreover, under the assumption of a large field - so large to jeopardize perturbation theory - the Fossils' approach is not justified (as one is not in the position to truncate the series) and one is forced to resort to this resummation technique. In this sense this technique literally extends the Fossils' approach, as it is justified also in a region of parameter space where the Fossils' approach is not. 
    \item The applicability of the non-perturbative technique hinges on the presence of a \textit{large} fluctuation in one of the fields. Classically this just means that a field is, for some physical reason, enhanced with respect to its (locally observable) average value, and being the non-perturbative technique semiclassical, this is enough. We remark that in any case this is a situation that may happen, accounting for the fact that the Fossil field has fluctuations on much larger scales than our observable Hubble patch. To match with the Fossils' approach, which in its later forms has been formalized in a in-in (quantum) field theoretic approach \cite{Dimastrogiovanni:2022afr}, a bit more interpretation is needed. On a QFT perspective, we can obtain large fluctuations considering the field operators to be applied to states that are coherent states with a very large number of particles \cite{Ragavendra:2024qpj, bialynicki2013quantum}: we are therefore considering not a high energy limit, but a high occupation number limit. 
    
    \item The benefit of the technique is not just its capability to bypass lengthy computations in perturbation theory (we remark that every interacting vertex is an integral in the in-in formalism in FLRW spacetime), but also the fact that it could unveil, phenomenologically speaking, genuinely new scenarios. Looking at the application of the technique to the concrete model of Inflation, for example, a modification of the speed of sound for the scalar $\zeta$ field is a truly new behavior, that could not have been gauged just from perturbation theory within the Fossils' prescription. This modification of the speed of sound, which could be significant in particular kinematical scenarios, is relevant for the enhancement of non-Gaussianities \cite{Chen_2007, Emery:2012sm, Bartolo_2004, Chen_2010}, and the production of Primordial Black Holes \cite{Matarrese:1986et, Lucchin:1987yv,Catelan:1988zb, Franciolini:2018vbk, Romano:2020gtn}. More in particular, this non-perturbative technique seems amenable to be combined with other non-perturbative approaches, such as the saddle-point integration or the cluster expansion, adopted in \cite{Matarrese:1986et, Lucchin:1987yv,Catelan:1988zb, Franciolini:2018vbk}. We leave these interesting explorations to future works. 
\end{itemize}

\acknowledgments
We thank E. Dimastrogiovanni and M. Fasiello for reading the draft. NB and RI wish to thank S. Matarrese for many useful and stimulating discussions in the early stages of this work. RI wishes to thank D. Comelli, M. Ballardini, L. Caloni and N. Barbieri for interesting discussions. RI is thankful to M. Braglia for interesting discussions and clarifications about field redefinitions. RI wishes to thank all the participants and the organizers of the workshop \vg The Dawn of GW Cosmology'', where this work was presented, for the stimulating conversations. NB and RI acknowledge the financial support of the INFN InDark initiative. NB acknowledges the financial support from the COSMOS network (www.cosmosnet.it) through the ASI (Italian Space
Agency) Grants 2016-24-H.0, 2016-24-H.1-2018 and
2020-9-HH.0. NB acknowledges support by the MUR PRIN2022 Project “BROWSEPOL: Beyond standaRd mOdel With coSmic microwavE background POLarization”-2022EJNZ53 financed by the European Union - Next Generation EU. 
\appendix
\section{Equations of motion}
In this appendix we write down for clarity the derivation of the equation of motion of a scalar field in de Sitter space. Dots $\dot{}$ denote derivatives with respect to comoving time $t$. 
The standard action of a massive scalar field in de Sitter space is \begin{equation}
    S=\int d^4\xx \sqrt{-g} \mathcal{L}=\int d^4\xx -\frac{1}{2} \sqrt{-g}\left(\partial_\mu \phi \partial_\nu \phi g^{\mu \nu}+ m^2 \phi^2\right),
\end{equation}
where in FLRW $ds^2=-dt^2+a(t)^2 d\xx^2=a(\eta)^2(-d\eta^2+d\xx^2)$, hence $\sqrt{-g}=a(t)^3=a(\eta)^4$. We employ the fact that, for a tensor $T^\mu$: 
\begin{equation}
    \nabla_\mu T^\mu = \frac{1}{\sqrt{-g}}\partial_\mu (\sqrt{-g}T^\mu). 
\end{equation}
The equation of motion read from first principles \begin{equation}
    \frac{\partial \mathcal{L}}{\partial \phi }=\nabla_\mu \left(\frac{\partial \mathcal{L}}{\partial \partial_\mu \phi}\right).
\end{equation}
We now compute pedantically chunk by chunk the equation, beginning with the left hand side (LHS)
\begin{equation}
    \frac{\partial \mathcal{L}}{\partial \phi }=-m^2\phi. 
    \end{equation}
    The right hand side (RHS) gives: 
    \begin{align}
     \nabla_\mu(\partial_\nu \phi g^{\mu \nu}) \notag 
    = \frac{1}{a^3} \partial_\mu (a^3 \partial ^\mu \phi )  \notag\\
    = 3\frac{\dot{a}}{a} \partial_0 \phi + \partial_\mu \partial^\mu \phi  \notag\\
    = -3H\dot{\phi} - \ddot{\phi}+ \frac{(\partial_i)^2}{a^2}\phi. 
    \end{align}
Combining the two we get
    \begin{equation}
        3H\dot{\phi} + \ddot{\phi}- \frac{(\partial_i)^2}{a^2}\phi - m^2\phi =0.
    \end{equation}
The same applies in conformal time, where $a(\eta)=-1/(H\eta)$: \begin{align}
    S = \int d\eta d^3 \xx \sqrt{-g} \mathcal{L}= \int d\eta d^3 \xx -\frac{1}{2}\sqrt{-g}(\partial_\mu \phi \partial_\nu \phi g^{\mu \nu }+ m^2\phi^2)  \notag\\ =  \int d\eta d^3 \xx \frac{-1}{2\eta^4H^4}\left(({H^2\eta^2})(\phi'\phi'-\partial_i\phi\partial_i\phi)+ {m^2\phi^2}\right).
\end{align}
The interesting part happens with the kinetic term, that reads: \begin{equation}
    \eta^4H^4\partial_\mu \left(\frac{-1}{\eta^2 H^2}(-\phi,\partial_i \phi )\right) = \eta^2 H^2 (\phi'' - \partial_i^2 \phi -\frac{2}{\eta}\phi'), 
\end{equation} making the equation of motion \begin{equation}
    0=\phi''-\frac{2}{\eta}\phi'-\partial_i^2\phi + \frac{m^2\phi}{\eta^2 H^2}. 
\end{equation}
One can check that the two correspond. We remind that \begin{equation}
    H(t)=\frac{1}{a(t) }\frac{da(t)}{dt}
\end{equation}while \begin{equation}
    H(\eta)= \frac{1}{a(\eta) }\frac{da(\eta)}{d\eta}
\end{equation} and since $dt = a d\eta$, one has $a(\eta)= -1/(\eta H) $.
\section{Useful integrals }
The following are useful integrals when dealing with the in-in formalism, we write them for clarity. We remark that the limit to $-\infty$ is tilted in the imaginary direction to ensure convergence. 
\begin{equation}
    \int_{-\infty }^{\eta'\to 0} d\eta e^{ik\eta}\eta = \frac{e^{ik\eta}}{k^2}\bigg{|}_{-\infty}^{\eta'}+\frac{\eta e^{ik\eta}}{-ik}\bigg{|}_{-\infty}^{\eta'}
\end{equation}
\begin{equation}
    \int_{-\infty }^{\eta'\to 0} d\eta e^{ik\eta}= \frac{e^{ik\eta}}{-ik}\bigg{|}_{-\infty}^{\eta'}
\end{equation}
\begin{equation}
    \int_{-\infty }^{\eta'\to 0} d\eta \frac{e^{-ik\eta}}{\eta}= \gamma_E+\text{log}(-i\eta' k) 
\end{equation}
\begin{equation}
    \int_{-\infty}^{\eta'\to 0} d\eta \frac{e^{-ik\eta}}{\eta^2} = \frac{-e^{-ik\eta}}{\eta}\bigg{|}_{-\infty}^{\eta'}-ik(\gamma_E+\text{log}(-i\eta' k) )
\end{equation}
\begin{equation}
    \int_{-\infty}^{\eta'\to 0} d\eta \frac{e^{-ik\eta}}{\eta^3} = -\frac{1}{2}\frac{e^{-ik\eta}}{\eta^2}\bigg{|}_{-\infty}^{\eta'}+\frac{ik}{2}\frac{-e^{-ik\eta}}{\eta}\bigg{|}_{-\infty}^{\eta'}-\frac{k^2}{2}(\gamma_E+\text{log}(-i\eta' k) )
\end{equation}
\begin{equation}
    \int_{-\infty}^{\eta'\to 0} d\eta \frac{e^{-ik\eta}}{\eta^4} = \frac{-1}{3}\frac{e^{-ik\eta}}{\eta^3}\bigg{|}_{-\infty}^{\eta'}+\frac{ik}{6}\frac{e^{-ik\eta}}{\eta^2}\bigg{|}_{-\infty}^{\eta'}+\frac{k^2}{6}\frac{-e^{-ik\eta}}{\eta}\bigg{|}_{-\infty}^{\eta'}+\frac{ik^3}{6}(\gamma_E+\text{log}(-i\eta' k) )
\end{equation}
\section{Concrete model: auxiliary field}
In this Appendix we provide the computations for the lengthy equation of motion for single field slow roll Inflation in the case of two gravitons and a scalar fluctuation. 
\newline 
The starting action reads: \begin{align}
    S=\int d^3\xx d\eta \frac{-\epsilon}{\eta^4H^4}\left[\zeta'\zeta'g^{00}+\partial_i\zeta \partial_j\zeta g^{ij}\right]-\frac{1}{8\eta^4H^4}\left[\gamma_{ij}'\gamma_{ij}'g^{00}+\partial_k\gamma_{ij}\partial_h\gamma_{ij}g^{hk}\right] \notag \\
    \frac{-\epsilon}{8\eta^4H^4}\left[\zeta \gamma_{ij}'\gamma_{ij}'g^{00}-\zeta\partial_k\gamma_{ij}\partial_h\gamma_{ij}g^{hk}\right]-\frac{1}{4\eta^4H^4}g^{hk}\left[\gamma_{ij}'\partial_k\gamma_{ij}{\partial_h}\chi\right].
\end{align}
The on-shell terms for the auxiliary field reads (in real and momentum space): \begin{equation}
    \nabla^2 \chi = \epsilon \zeta', \; \; q^2\chi = \epsilon \zeta',
\end{equation} and therefore they are of the same order in slow roll as the other interacting term. 
We derive the $\nabla_\mu(\partial \mathcal{L}/\partial(\partial_\mu \gamma))$ terms: 
\begin{align}
    \frac{\eta^2 H^2}{4}\left(\gamma_{ij}''-\frac{2}{\eta}\gamma_{ij}'-\nabla^2\gamma_{ij}\right) 
    + \frac{\epsilon \eta^2H^2}{4}\left(\zeta'\gamma_{ij}'+\gamma_{ij}''\zeta -\frac{2}{\eta}\zeta \gamma_{ij}'+\partial_h\zeta \partial_h\gamma_{ij}+\zeta \nabla^2 \gamma_{ij} \right) \notag \\
    + \frac{ \eta^2H^2}{4}\bigg[-\frac{2}{\eta}\left(\partial_h\chi \right)\partial_h\gamma_{ij}+ \left(\partial_h\chi'\right)\partial_h\gamma_{ij}+ 2\left(\partial_h\chi\right)\partial_h\gamma_{ij}'+\epsilon \zeta' \gamma_{ij}'\bigg]=0. 
\end{align} In the above equation the first line comes respectively from the kinetic term and the first interacting term in the action, the last line instead is the chunk that comes from the last interacting term in the action. As we move to momentum space and get rid of overall factors, we get the long but a bit clearer equation: \begin{align}
 \gamma''_{ij}(\kk)-\frac{2}{\eta}\gamma'_{ij}(\kk)+k^2\gamma_{ij}(\kk)
   +\epsilon\bigg[\int \frac{d^3\qq}{(2\pi)^3}\bigg(\zeta'(\qq)\gamma_{ij}'(\kk-\qq)+\zeta(\qq) \gamma_{ij}''(\kk-\qq)- \notag \\ \frac{2}{\eta}\zeta(\qq)\gamma_{ij}'(\kk-\qq)-\qq\cdot(\kk-\qq)\zeta(\qq)\gamma_{ij}(\kk-\qq)-(\kk-\qq)^2\zeta(\qq)\gamma_{ij}(\kk-\qq)\bigg)\bigg] \notag \\
   + \int \frac{d^3\qq}{(2\pi)^3} \bigg[\bigg( -\frac{2}{\eta} {\qq\cdot(\kk-\qq)}\chi(\qq)\gamma_{ij}(\kk-\qq)+{\qq\cdot(\kk-\qq)}\chi'(\qq)\gamma_{ij}(\kk- \qq)  \notag \\
   -2{\qq\cdot(\kk-\qq)}\chi(\qq)\gamma_{ij}'(\kk -\qq)+\epsilon\zeta'(\qq)\gamma_{ij}'(\kk-\qq)\bigg)\bigg]=0. 
\end{align}
In the above equation again the last two lines are the result of the last line of the equation of motion in real space. Substituting in the last line the definition of $\chi$ on shell, we reduce the last line to 
\begin{equation}
    \epsilon \int \frac{d^3\qq}{(2\pi)^3} \bigg[\bigg( \frac{2}{\eta} \frac{\qq\cdot(\kk-\qq)}{q^2}\zeta'\gamma_{ij}- \frac{\qq\cdot(\kk-\qq)}{q^2}\zeta''\gamma_{ij}-2\frac{\qq\cdot(\kk-\qq)}{q^2}\zeta'\gamma_{ij}'-\zeta'\gamma_{ij}'\bigg)\bigg]. 
\end{equation}
We compute now the equation deriving with respect to the constrain: \begin{equation}
    0= \partial_h\gamma_{ij}'\partial_h\gamma_{ij}+\gamma_{ij}'\nabla^2\gamma_{ij},
\end{equation}and the equation for $\zeta$: 
\begin{equation}
    \zeta''(\qq)-\frac{2}{\eta}\zeta'(\qq)+k^2\zeta(\qq)=-\frac{1}{H}\int \frac{d^3\kk}{(2\pi)^3}\left(\kk\cdot(\qq-\kk)\gamma_{ij}(\kk)\gamma_{ij}(\qq-\kk)-\gamma_{ij}(\kk)'\gamma_{ij}(\qq-\kk)'\right).
\end{equation}We work in the assumption in which the mode of $\zeta$ is much larger than the mode of $\gamma$, therefore we neglect the RHS of the $\zeta$ equation, which solution (and its derivatives) read as usual: \begin{align}
    \zeta(\qq)=(2\pi)^3\delta^{(3)}(\qq-\kk_L)\bar{\zeta}(1-iq\eta)e^{iq\eta} \notag \\
    \zeta'(\qq)=(2\pi)^3\delta^{(3)}(\qq-\kk_L)\bar{\zeta}\eta q^2 e^{iq\eta}  \notag \\
    \zeta''(\qq)=(2\pi)^3\delta^{(3)}(\qq-\kk_L)\bar{\zeta}(q^2+iq^3\eta)e^{iq\eta}. 
\end{align}
We are also fixing the kinematics of the triangle in momentum space to be an isosceles triangle with short side $\kk_L$ and long (equal) sides $\kk$, and we therefore remark that, as $\kk_L\to0$ \begin{equation}
    \kk_L\kk\sim -\frac{1}{2}k_L^2,
\end{equation} for the law of cosines. The equation that comes from the constrain does not give us valuable information. Plugging the solution of $\zeta$ in the equation for $\gamma$, we get the lengthy but clear (omitting the momentum dependencies): 
\begin{align}
&\gamma''_{ij}-\frac{2}{\eta}\gamma_{ij}'+ k^2\gamma_{ij} \notag 
+ \epsilon\bigg[\bar{\zeta}e^{ik_L\eta}k_L^2\eta\gamma_{ij}'+\bar{\zeta}(1-ik_L\eta)e^{ik_L\eta}\gamma_{ij}''\\ &-\frac{2}{\eta}\bar{\zeta}(1-ik_L\eta)e^{ik_L\eta}\gamma_{ij}'\notag  -\kk_L\kk \bar{\zeta}(1-ik_L\eta)e^{ik_L\eta}\gamma_{ij}-k^2\bar{\zeta}(1-ik_L\eta)e^{ik_L\eta}\gamma\bigg] \notag \\&
    + \epsilon\bigg[-\gamma_{ij}k_L^2\bar{\zeta}e^{ik_L\eta}+\gamma_{ij}(k_L^2+ik_L^3\eta)e^{ik_L\eta}\bar{\zeta}+2k_L^2\eta e^{ik_L\eta} \bar{\zeta}\gamma_{ij}'+k_L^2\eta e^{ik_L\eta} \bar{\zeta}\gamma_{ij}'\bigg]=0.
\end{align}
In the limit $\eta \to 0, \kk_L\to 0$ the only surviving terms are \begin{equation}
    0= \left(\gamma''_{ij}-\frac{2}{\eta}\gamma_{ij}'\right)(1+\epsilon \bar{\zeta})+k^2\gamma_{ij}(1-\epsilon \bar{\zeta}). 
\end{equation}
This last interesting equation is the equation of motion of the field $\gamma$ after we integrated out the now auxiliary degree of freedom $\zeta$, and that we used in the non-perturbative computation in the bulk of the paper, namely Eq. (\ref{eom zeta}). 

\bibliographystyle{JHEP}
\bibliography{biblio}

@article{Bartolo_2004,
	title        = {Non-Gaussianity from inflation: theory and observations},
	author       = {N. Bartolo and E. Komatsu and S. Matarrese and A. Riotto},
	year         = 2004,
	month        = {nov},
	journal      = {Physics Reports},
	publisher    = {Elsevier {BV}},
	volume       = 402,
	number       = {3-4},
	pages        = {103--266},
	doi          = {10.1016/j.physrep.2004.08.022},
	url          = {https://doi.org/10.1016%2Fj.physrep.2004.08.022}
}

@article{Chen_2010,
	title        = {Primordial Non-Gaussianities from Inflation Models},
	author       = {Xingang Chen},
	year         = 2010,
	journal      = {Advances in Astronomy},
	publisher    = {Hindawi Limited},
	volume       = 2010,
	pages        = {1--43},
	doi          = {10.1155/2010/638979},
	url          = {https://doi.org/10.1155%2F2010%2F638979}
}

@article{Chen_2017,
	title        = {Schwinger-Keldysh diagrammatics for primordial perturbations},
	author       = {Xingang Chen and Yi Wang and Zhong-Zhi Xianyu},
	year         = 2017,
	month        = {dec},
	journal      = {Journal of Cosmology and Astroparticle Physics},
	publisher    = {{IOP} Publishing},
	volume       = 2017,
	number       = 12,
	pages        = {006--006},
	doi          = {10.1088/1475-7516/2017/12/006},
	url          = {https://doi.org/10.1088%2F1475-7516%2F2017%2F12%2F006}
}

@article{Ragavendra:2024qpj,
    author = "Ragavendra, H. V. and Mukherjee, Dipayan and Sethi, Shiv K.",
    title = "{Cosmological consequences of statistical inhomogeneity}",
    eprint = "2411.01331",
    archivePrefix = "arXiv",
    primaryClass = "astro-ph.CO",
    month = "11",
    year = "2024"
}

@article{Emery:2012sm,
    author = "Emery, Jon and Tasinato, Gianmassimo and Wands, David",
    title = "{Local non-Gaussianity from rapidly varying sound speeds}",
    eprint = "1203.6625",
    archivePrefix = "arXiv",
    primaryClass = "hep-th",
    doi = "10.1088/1475-7516/2012/08/005",
    journal = "JCAP",
    volume = "08",
    pages = "005",
    year = "2012"
}

@article{Seery_2008,
   title={Non-Gaussianity of inflationary field perturbations from the field equation},
   volume={2008},
   ISSN={1475-7516},
   url={http://dx.doi.org/10.1088/1475-7516/2008/03/014},
   DOI={10.1088/1475-7516/2008/03/014},
   number={03},
   journal={Journal of Cosmology and Astroparticle Physics},
   publisher={IOP Publishing},
   author={Seery, David and Malik, Karim A and Lyth, David H},
   year={2008},
   month=mar, pages={014} }

@article{Lyth_2005_ng,
   title={Inflationary Prediction for Primordial Non-Gaussianity},
   volume={95},
   ISSN={1079-7114},
   url={http://dx.doi.org/10.1103/PhysRevLett.95.121302},
   DOI={10.1103/physrevlett.95.121302},
   number={12},
   journal={Physical Review Letters},
   publisher={American Physical Society (APS)},
   author={Lyth, David H. and Rodríguez, Yeinzon},
   year={2005},
   month=sep }

@article{Lyth_2005,
   title={A general proof of the conservation of the curvature perturbation},
   volume={2005},
   ISSN={1475-7516},
   url={http://dx.doi.org/10.1088/1475-7516/2005/05/004},
   DOI={10.1088/1475-7516/2005/05/004},
   number={05},
   journal={Journal of Cosmology and Astroparticle Physics},
   publisher={IOP Publishing},
   author={Lyth, David H and Malik, Karim A and Sasaki, Misao},
   year={2005},
   month=may, pages={004–004} }

@article{Gangui_1994,
   title={The three-point correlation function of the cosmic microwave background in inflationary models},
   volume={430},
   ISSN={1538-4357},
   url={http://dx.doi.org/10.1086/174421},
   DOI={10.1086/174421},
   journal={The Astrophysical Journal},
   publisher={American Astronomical Society},
   author={Gangui, Alejandro and Lucchin, Francesco and Matarrese, Sabino and Mollerach, Silvia},
   year={1994},
   month=aug, pages={447} }

@article{Weinberg_2005,
	title        = {Quantum contributions to cosmological correlations},
	author       = {Steven Weinberg},
	year         = 2005,
	month        = {aug},
	journal      = {Physical Review D},
	publisher    = {American Physical Society ({APS})},
	volume       = 72,
	number       = 4,
	doi          = {10.1103/physrevd.72.043514},
	url          = {https://doi.org/10.1103%2Fphysrevd.72.043514}
}

@article{Huenupi_2024,
   title={Regularizing infrared divergences in de Sitter spacetime: Loops, dimensional regularization, and cutoffs},
   volume={110},
   ISSN={2470-0029},
   url={http://dx.doi.org/10.1103/PhysRevD.110.123536},
   DOI={10.1103/physrevd.110.123536},
   number={12},
   journal={Physical Review D},
   publisher={American Physical Society (APS)},
   author={Huenupi, Javier and Hughes, Ellie and Palma, Gonzalo A. and Sypsas, Spyros},
   year={2024},
   month={dec},
}

@article{Palma_2025,
   title={Non-Gaussian statistics of de Sitter spectators: a perturbative derivation of stochastic dynamics},
   volume={2025},
   ISSN={1029-8479},
   url={http://dx.doi.org/10.1007/JHEP12(2025)170},
   DOI={10.1007/jhep12(2025)170},
   number={12},
   journal={Journal of High Energy Physics},
   publisher={Springer Science and Business Media LLC},
   author={Palma, Gonzalo A. and Sypsas, Spyros},
   year={2025},
   month=dec }

@article{Braglia_2024,
   title={No time to derive: unraveling total time derivatives in in-in perturbation theory},
   volume={2024},
   ISSN={1029-8479},
   url={http://dx.doi.org/10.1007/JHEP08(2024)068},
   DOI={10.1007/jhep08(2024)068},
   number={8},
   journal={Journal of High Energy Physics},
   publisher={Springer Science and Business Media LLC},
   author={Braglia, Matteo and Pinol, Lucas},
   year={2024},
   month=aug }

@article{Arroja_2011,
   title={A note on the role of the boundary terms for the non-Gaussianity in general k-inflation},
   volume={2011},
   ISSN={1475-7516},
   url={http://dx.doi.org/10.1088/1475-7516/2011/05/005},
   DOI={10.1088/1475-7516/2011/05/005},
   number={05},
   journal={Journal of Cosmology and Astroparticle Physics},
   publisher={IOP Publishing},
   author={Arroja, Frederico and Tanaka, Takahiro},
   year={2011},
   month=may, pages={005–005} }

@article{Riotto_2008,
   title={On resumming inflationary perturbations beyond one-loop},
   volume={2008},
   ISSN={1475-7516},
   url={http://dx.doi.org/10.1088/1475-7516/2008/04/030},
   DOI={10.1088/1475-7516/2008/04/030},
   number={04},
   journal={Journal of Cosmology and Astroparticle Physics},
   publisher={IOP Publishing},
   author={Riotto, Antonio and Sloth, Martin S},
   year={2008},
   month=apr, pages={030} }

@misc{calabrese2025atacamacosmologytelescopedr6,
      title={The Atacama Cosmology Telescope: DR6 Constraints on Extended Cosmological Models}, 
      author={Erminia Calabrese and J. Colin Hill and Hidde T. Jense and Adrien La Posta and Irene Abril-Cabezas and Graeme E. Addison and Peter A. R. Ade and Simone Aiola and Tommy Alford and David Alonso and Mandana Amiri and Rui An and Zachary Atkins and Jason E. Austermann and Eleonora Barbavara and Nicola Barbieri and Nicholas Battaglia and Elia Stefano Battistelli and James A. Beall and Rachel Bean and Ali Beheshti and Benjamin Beringue and Tanay Bhandarkar and Emily Biermann and Boris Bolliet and J Richard Bond and Valentina Capalbo and Felipe Carrero and Shi-Fan Chen and Grace Chesmore and Hsiao-mei Cho and Steve K. Choi and Susan E. Clark and Nicholas F. Cothard and Kevin Coughlin and William Coulton and Devin Crichton and Kevin T. Crowley and Omar Darwish and Mark J. Devlin and Simon Dicker and Cody J. Duell and Shannon M. Duff and Adriaan J. Duivenvoorden and Jo Dunkley and Rolando Dunner and Carmen Embil Villagra and Max Fankhanel and Gerrit S. Farren and Simone Ferraro and Allen Foster and Rodrigo Freundt and Brittany Fuzia and Patricio A. Gallardo and Xavier Garrido and Martina Gerbino and Serena Giardiello and Ajay Gill and Jahmour Givans and Vera Gluscevic and Samuel Goldstein and Joseph E. Golec and Yulin Gong and Yilun Guan and Mark Halpern and Ian Harrison and Matthew Hasselfield and Adam He and Erin Healy and Shawn Henderson and Brandon Hensley and Carlos Hervías-Caimapo and Gene C. Hilton and Matt Hilton and Adam D. Hincks and Renée Hložek and Shuay-Pwu Patty Ho and John Hood and Erika Hornecker and Zachary B. Huber and Johannes Hubmayr and Kevin M. Huffenberger and John P. Hughes and Margaret Ikape and Kent Irwin and Giovanni Isopi and Neha Joshi and Ben Keller and Joshua Kim and Kenda Knowles and Brian J. Koopman and Arthur Kosowsky and Darby Kramer and Aleksandra Kusiak and Alex Lague and Victoria Lakey and Massimiliano Lattanzi and Eunseong Lee and Yaqiong Li and Zack Li and Michele Limon and Martine Lokken and Thibaut Louis and Marius Lungu and Niall MacCrann and Amanda MacInnis and Mathew S. Madhavacheril and Diego Maldonado and Felipe Maldonado and Maya Mallaby-Kay and Gabriela A. Marques and Joshiwa van Marrewijk and Fiona McCarthy and Jeff McMahon and Yogesh Mehta and Felipe Menanteau and Kavilan Moodley and Thomas W. Morris and Tony Mroczkowski and Sigurd Naess and Toshiya Namikawa and Federico Nati and Simran K. Nerval and Laura Newburgh and Andrina Nicola and Michael D. Niemack and Michael R. Nolta and John Orlowski-Scherer and Luca Pagano and Lyman A. Page and Shivam Pandey and Bruce Partridge and Karen Perez Sarmiento and Heather Prince and Roberto Puddu and Frank J. Qu and Damien C. Ragavan and Bernardita Ried Guachalla and Keir K. Rogers and Felipe Rojas and Tai Sakuma and Emmanuel Schaan and Benjamin L. Schmitt and Neelima Sehgal and Shabbir Shaikh and Blake D. Sherwin and Carlos Sierra and Jon Sievers and Cristóbal Sifón and Sara Simon and Rita Sonka and David N. Spergel and Suzanne T. Staggs and Emilie Storer and Kristen Surrao and Eric R. Switzer and Niklas Tampier and Leander Thiele and Robert Thornton and Hy Trac and Carole Tucker and Joel Ullom and Leila R. Vale and Alexander Van Engelen and Jeff Van Lanen and Cristian Vargas and Eve M. Vavagiakis and Kasey Wagoner and Yuhan Wang and Lukas Wenzl and Edward J. Wollack and Kaiwen Zheng},
      year={2025},
      eprint={2503.14454},
      archivePrefix={arXiv},
      primaryClass={astro-ph.CO},
      url={https://arxiv.org/abs/2503.14454}, 
}

@article{Celoria_2021,
	title        = {Beyond perturbation theory in inflation},
	author       = {Marco Celoria and Paolo Creminelli and Giovanni Tambalo and Vicharit Yingcharoenrat},
	year         = 2021,
	month        = {jun},
	journal      = {Journal of Cosmology and Astroparticle Physics},
	publisher    = {{IOP} Publishing},
	volume       = 2021,
	number       = {06},
	pages        = {051},
	doi          = {10.1088/1475-7516/2021/06/051},
	url          = {https://doi.org/10.1088%2F1475-7516%2F2021%2F06%2F051}
}

@article{Starobinsky:1980te,
	title        = {{A New Type of Isotropic Cosmological Models Without Singularity}},
	author       = {Starobinsky, Alexei A.},
	year         = 1980,
	journal      = {Phys. Lett. B},
	volume       = 91,
	pages        = {99--102},
	doi          = {10.1016/0370-2693(80)90670-X},
	editor       = {Khalatnikov, I. M. and Mineev, V. P.}
}

@article{Guth:1980zm,
	title        = {{The Inflationary Universe: A Possible Solution to the Horizon and Flatness Problems}},
	author       = {Guth, Alan H.},
	year         = 1981,
	journal      = {Phys. Rev. D},
	volume       = 23,
	pages        = {347--356},
	doi          = {10.1103/PhysRevD.23.347},
	editor       = {Fang, Li-Zhi and Ruffini, R.},
	reportnumber = {SLAC-PUB-2576}
}

@article{Linde:1981mu,
	title        = {{A New Inflationary Universe Scenario: A Possible Solution of the Horizon, Flatness, Homogeneity, Isotropy and Primordial Monopole Problems}},
	author       = {Linde, Andrei D.},
	year         = 1982,
	journal      = {Phys. Lett. B},
	volume       = 108,
	pages        = {389--393},
	doi          = {10.1016/0370-2693(82)91219-9},
	editor       = {Fang, Li-Zhi and Ruffini, R.},
	reportnumber = {LEBEDEV-81-229}
}

@article{Ach_carro_2022,
	title        = {The hand-made tail: non-perturbative tails from multifield inflation},
	author       = {Ana Ach{\'{u}}carro and Sebasti{\'{a}}n C{\'{e}}spedes and Anne-Christine Davis and Gonzalo A. Palma},
	year         = 2022,
	month        = {may},
	journal      = {Journal of High Energy Physics},
	publisher    = {Springer Science and Business Media {LLC}},
	volume       = 2022,
	number       = 5,
	doi          = {10.1007/jhep05(2022)052},
	url          = {https://doi.org/10.1007%2Fjhep05%282022%29052}
}

@article{Planck:2018jri,
	title        = {{Planck 2018 results. X. Constraints on inflation}},
	author       = {Akrami, Y. and others},
	year         = 2020,
	journal      = {Astron. Astrophys.},
	volume       = 641,
	pages        = {A10},
	doi          = {10.1051/0004-6361/201833887},
	collaboration = {Planck},
	eprint       = {1807.06211},
	archiveprefix = {arXiv},
	primaryclass = {astro-ph.CO}
}

@article{Cheung_2008,
	title        = {The effective field theory of inflation},
	author       = {Clifford Cheung and A. Liam Fitzpatrick and Jared Kaplan and Leonardo Senatore and Paolo Creminelli},
	year         = 2008,
	month        = {mar},
	journal      = {Journal of High Energy Physics},
	publisher    = {Springer Science and Business Media {LLC}},
	volume       = 2008,
	number       = {03},
	pages        = {014--014},
	doi          = {10.1088/1126-6708/2008/03/014},
	url          = {https://doi.org/10.1088%2F1126-6708%2F2008%2F03%2F014}
}

@article{Weinberg_2008,
	title        = {Effective field theory for inflation},
	author       = {Steven Weinberg},
	year         = 2008,
	month        = {jun},
	journal      = {Physical Review D},
	publisher    = {American Physical Society ({APS})},
	volume       = 77,
	number       = 12,
	doi          = {10.1103/physrevd.77.123541},
	url          = {https://doi.org/10.1103%2Fphysrevd.77.123541}
}

@article{Maldacena_2003,
	title        = {Non-gaussian features of primordial fluctuations in single field inflationary models},
	author       = {Juan Maldacena},
	year         = 2003,
	month        = {may},
	journal      = {Journal of High Energy Physics},
	publisher    = {Springer Science and Business Media {LLC}},
	volume       = 2003,
	number       = {05},
	pages        = {013--013},
	doi          = {10.1088/1126-6708/2003/05/013},
	url          = {https://doi.org/10.1088%2F1126-6708%2F2003%2F05%2F013}
}

@article{Catelan:1988zb,
	title        = {{Peak Number Density of Nongaussian Random Fields}},
	author       = {Catelan, Paolo and Lucchin, Francesco and Matarrese, Sabino},
	year         = 1988,
	journal      = {Phys. Rev. Lett.},
	volume       = 61,
	pages        = {267--270},
	doi          = {10.1103/PhysRevLett.61.267},
	reportnumber = {DFPD-12-88}
}

@article{Matarrese:1986et,
	title        = {{A PATH INTEGRAL APPROACH TO LARGE SCALE MATTER DISTRIBUTION ORIGINATED BY NONGAUSSIAN FLUCTUATIONS}},
	author       = {Matarrese, Sabino and Lucchin, Francesco and Bonometto, Silvio A.},
	year         = 1986,
	journal      = {Astrophys. J. Lett.},
	volume       = 310,
	pages        = {L21--L26},
	doi          = {10.1086/184774},
	reportnumber = {SISSA-41-86-EP}
}

@article{Lucchin:1987yv,
	title        = {{The Effect of nonGaussian statistics on the mass multiplicity of cosmic structures}},
	author       = {Lucchin, Francesco and Matarrese, Sabino},
	year         = 1988,
	journal      = {Astrophys. J.},
	volume       = 330,
	pages        = {535--544},
	doi          = {10.1086/166492},
	reportnumber = {DFPD-33-87}
}

@article{Romano:2020gtn,
    author = "Romano, Antonio Enea",
    title = "{Sound speed induced production of primordial black holes}",
    eprint = "2006.07321",
    archivePrefix = "arXiv",
    primaryClass = "astro-ph.CO",
    month = "6",
    year = "2020"
}

@article{Chen_2007,
   title={Observational signatures and non-Gaussianities of general single-field inflation},
   volume={2007},
   ISSN={1475-7516},
   url={http://dx.doi.org/10.1088/1475-7516/2007/01/002},
   DOI={10.1088/1475-7516/2007/01/002},
   number={01},
   journal={Journal of Cosmology and Astroparticle Physics},
   publisher={IOP Publishing},
   author={Chen, Xingang and Huang, Min-xin and Kachru, Shamit and Shiu, Gary},
   year={2007},
   month=jan, pages={002–002} }

@article{Werth:2023pfl,
    author = "Werth, Denis and Pinol, Lucas and Renaux-Petel, S\'ebastien",
    title = "{Cosmological Flow of Primordial Correlators}",
    eprint = "2302.00655",
    archivePrefix = "arXiv",
    primaryClass = "hep-th",
    doi = "10.1103/PhysRevLett.133.141002",
    journal = "Phys. Rev. Lett.",
    volume = "133",
    number = "14",
    pages = "141002",
    year = "2024"
}

@article{Pinol:2023oux,
    author = "Pinol, Lucas and Renaux-Petel, S\'ebastien and Werth, Denis",
    title = "{The Cosmological Flow: A Systematic Approach to Primordial Correlators}",
    eprint = "2312.06559",
    archivePrefix = "arXiv",
    primaryClass = "astro-ph.CO",
    month = "12",
    year = "2023"
}

@article{Franciolini:2018vbk,
	title        = {{Primordial Black Holes from Inflation and non-Gaussianity}},
	author       = {Franciolini, G. and Kehagias, A. and Matarrese, S. and Riotto, A.},
	year         = 2018,
	journal      = {JCAP},
	volume       = {03},
	pages        = {016},
	doi          = {10.1088/1475-7516/2018/03/016},
	eprint       = {1801.09415},
	archiveprefix = {arXiv},
	primaryclass = {astro-ph.CO}
}

@article{Rubin_2001,
   title={Effect of massive fields on inflation},
   volume={74},
   ISSN={1090-6487},
   url={http://dx.doi.org/10.1134/1.1417158},
   DOI={10.1134/1.1417158},
   number={5},
   journal={Journal of Experimental and Theoretical Physics Letters},
   publisher={Pleiades Publishing Ltd},
   author={Rubin, S. G.},
   year={2001},
   month=sep, pages={247–250} }

@article{Tolley:2009fg,
    author = "Tolley, Andrew J. and Wyman, Mark",
    title = "{The Gelaton Scenario: Equilateral non-Gaussianity from multi-field dynamics}",
    eprint = "0910.1853",
    archivePrefix = "arXiv",
    primaryClass = "hep-th",
    reportNumber = "PI-COSMO-153",
    doi = "10.1103/PhysRevD.81.043502",
    journal = "Phys. Rev. D",
    volume = "81",
    pages = "043502",
    year = "2010"
}

@article{Dimastrogiovanni:2012st,
    author = "Dimastrogiovanni, Emanuela and Fasiello, Matteo and Tolley, Andrew J.",
    title = "{Low-Energy Effective Field Theory for Chromo-Natural Inflation}",
    eprint = "1211.1396",
    archivePrefix = "arXiv",
    primaryClass = "hep-th",
    doi = "10.1088/1475-7516/2013/02/046",
    journal = "JCAP",
    volume = "02",
    pages = "046",
    year = "2013"
}

@article{Guilleux:2016oqv,
    author = "Guilleux, Maxime and Serreau, Julien",
    title = "{Nonperturbative renormalization group for scalar fields in de Sitter space: beyond the local potential approximation}",
    eprint = "1611.08106",
    archivePrefix = "arXiv",
    primaryClass = "gr-qc",
    doi = "10.1103/PhysRevD.95.045003",
    journal = "Phys. Rev. D",
    volume = "95",
    number = "4",
    pages = "045003",
    year = "2017"
}

@article{Caravano_2025,
   title={Ultraslow-roll inflation on the lattice: Backreaction and nonlinear effects},
   volume={111},
   ISSN={2470-0029},
   url={http://dx.doi.org/10.1103/PhysRevD.111.063518},
   DOI={10.1103/physrevd.111.063518},
   number={6},
   journal={Physical Review D},
   publisher={American Physical Society (APS)},
   author={Caravano, Angelo and Franciolini, Gabriele and Renaux-Petel, Sébastien},
   year={2025},
   month=mar }

@article{Caravano_2024,
   title={Inflationary Butterfly Effect: Nonperturbative Dynamics from Small-Scale Features},
   volume={133},
   ISSN={1079-7114},
   url={http://dx.doi.org/10.1103/PhysRevLett.133.151001},
   DOI={10.1103/physrevlett.133.151001},
   number={15},
   journal={Physical Review Letters},
   publisher={American Physical Society (APS)},
   author={Caravano, Angelo and Inomata, Keisuke and Renaux-Petel, Sébastien},
   year={2024},
   month=oct }

@article{Caravano:2025diq,
    author = "Caravano, Angelo and Franciolini, Gabriele and Renaux-Petel, S{\'e}bastien",
    title = "{Ultra-Slow-Roll Inflation on the Lattice II: Nonperturbative Curvature Perturbation}",
    eprint = "2506.11795",
    archivePrefix = "arXiv",
    primaryClass = "astro-ph.CO",
    month = "6",
    year = "2025"
}

@article{Chen_2013,
   title={A single field inflation model with large local non-Gaussianity},
   volume={102},
   ISSN={1286-4854},
   url={http://dx.doi.org/10.1209/0295-5075/102/59001},
   DOI={10.1209/0295-5075/102/59001},
   number={5},
   journal={EPL (Europhysics Letters)},
   publisher={IOP Publishing},
   author={Chen, Xingang and Firouzjahi, Hassan and Hossein Namjoo, Mohammad and Sasaki, Misao},
   year={2013},
   month=jun, pages={59001} }

@article{Celoria_2017,
   title={Fluids, superfluids and supersolids: dynamics and cosmology of self-gravitating media},
   volume={2017},
   ISSN={1475-7516},
   url={http://dx.doi.org/10.1088/1475-7516/2017/09/036},
   DOI={10.1088/1475-7516/2017/09/036},
   number={09},
   journal={Journal of Cosmology and Astroparticle Physics},
   publisher={IOP Publishing},
   author={Celoria, Marco and Comelli, Denis and Pilo, Luigi},
   year={2017},
   month=sep, pages={036–036} }

@article{Celoria_2021_denis,
   title={Boosting GWs in supersolid inflation},
   volume={2021},
   ISSN={1029-8479},
   url={http://dx.doi.org/10.1007/JHEP01(2021)185},
   DOI={10.1007/jhep01(2021)185},
   number={1},
   journal={Journal of High Energy Physics},
   publisher={Springer Science and Business Media LLC},
   author={Celoria, Marco and Comelli, Denis and Pilo, Luigi and Rollo, Rocco},
   year={2021},
   month=jan }

@article{Weinberg_2003,
   title={Adiabatic modes in cosmology},
   volume={67},
   ISSN={1089-4918},
   url={http://dx.doi.org/10.1103/PhysRevD.67.123504},
   DOI={10.1103/physrevd.67.123504},
   number={12},
   journal={Physical Review D},
   publisher={American Physical Society (APS)},
   author={Weinberg, Steven},
   year={2003},
   month=jun }

@article{Wands_2000,
   title={New approach to the evolution of cosmological perturbations on large scales},
   volume={62},
   ISSN={1089-4918},
   url={http://dx.doi.org/10.1103/PhysRevD.62.043527},
   DOI={10.1103/physrevd.62.043527},
   number={4},
   journal={Physical Review D},
   publisher={American Physical Society (APS)},
   author={Wands, David and Malik, Karim A. and Lyth, David H. and Liddle, Andrew R.},
   year={2000},
   month=jul }

@article{Wands_2010,
   title={Local non-Gaussianity from inflation},
   volume={27},
   ISSN={1361-6382},
   url={http://dx.doi.org/10.1088/0264-9381/27/12/124002},
   DOI={10.1088/0264-9381/27/12/124002},
   number={12},
   journal={Classical and Quantum Gravity},
   publisher={IOP Publishing},
   author={Wands, David},
   year={2010},
   month=may, pages={124002} }

@article{Byrnes_2006,
   title={Primordial trispectrum from inflation},
   volume={74},
   ISSN={1550-2368},
   url={http://dx.doi.org/10.1103/PhysRevD.74.123519},
   DOI={10.1103/physrevd.74.123519},
   number={12},
   journal={Physical Review D},
   publisher={American Physical Society (APS)},
   author={Byrnes, Christian T. and Sasaki, Misao and Wands, David},
   year={2006},
   month=dec }

@article{Endlich_2013,
   title={Solid inflation},
   volume={2013},
   ISSN={1475-7516},
   url={http://dx.doi.org/10.1088/1475-7516/2013/10/011},
   DOI={10.1088/1475-7516/2013/10/011},
   number={10},
   journal={Journal of Cosmology and Astroparticle Physics},
   publisher={IOP Publishing},
   author={Endlich, Solomon and Nicolis, Alberto and Wang, Junpu},
   year={2013},
   month=oct, pages={011–011} }

@article{Fujita_2014,
   title={Non-perturbative approach for curvature perturbations in stochastic deltaNformalism},
   volume={2014},
   ISSN={1475-7516},
   url={http://dx.doi.org/10.1088/1475-7516/2014/10/030},
   DOI={10.1088/1475-7516/2014/10/030},
   number={10},
   journal={Journal of Cosmology and Astroparticle Physics},
   publisher={IOP Publishing},
   author={Fujita, Tomohiro and Kawasaki, Masahiro and Tada, Yuichiro},
   year={2014},
   month=oct, pages={030–030} }

@article{Fujita_2013,
   title={A new algorithm for calculating the curvature perturbations in stochastic inflation},
   volume={2013},
   ISSN={1475-7516},
   url={http://dx.doi.org/10.1088/1475-7516/2013/12/036},
   DOI={10.1088/1475-7516/2013/12/036},
   number={12},
   journal={Journal of Cosmology and Astroparticle Physics},
   publisher={IOP Publishing},
   author={Fujita, Tomohiro and Kawasaki, Masahiro and Tada, Yuichiro and Takesako, Tomohiro},
   year={2013},
   month=dec, pages={036–036} }

@misc{creminelli2024nonperturbativewavefunctionuniverseinflation,
      title={Non-perturbative Wavefunction of the Universe in Inflation with (Resonant) Features}, 
      author={Paolo Creminelli and Sébastien Renaux-Petel and Giovanni Tambalo and Vicharit Yingcharoenrat},
      year={2024},
      eprint={2401.10212},
      archivePrefix={arXiv},
      primaryClass={hep-th},
      url={https://arxiv.org/abs/2401.10212}, 
}

@article{Ach_carro_2012,
   title={Effective theories of single field inflation when heavy fields matter},
   volume={2012},
   ISSN={1029-8479},
   url={http://dx.doi.org/10.1007/JHEP05(2012)066},
   DOI={10.1007/jhep05(2012)066},
   number={5},
   journal={Journal of High Energy Physics},
   publisher={Springer Science and Business Media LLC},
   author={Achúcarro, Ana and Gong, Jinn-Ouk and Hardeman, Sjoerd and Palma, Gonzalo A. and Patil, Subodh P.},
   year={2012},
   month=may }

@article{Ach_carro_2015,
   title={On the viability ofm2$\phi$2and natural inflation},
   volume={2015},
   ISSN={1475-7516},
   url={http://dx.doi.org/10.1088/1475-7516/2015/07/008},
   DOI={10.1088/1475-7516/2015/07/008},
   number={07},
   journal={Journal of Cosmology and Astroparticle Physics},
   publisher={IOP Publishing},
   author={Achúcarro, Ana and Atal, Vicente and Welling, Yvette},
   year={2015},
   month=jul, pages={008–008} }

@article{Jeong:2012df,
	title        = {{Clustering Fossils from the Early Universe}},
	author       = {Jeong, Donghui and Kamionkowski, Marc},
	year         = 2012,
	journal      = {Phys. Rev. Lett.},
	volume       = 108,
	pages        = 251301,
	doi          = {10.1103/PhysRevLett.108.251301},
	eprint       = {1203.0302},
	archiveprefix = {arXiv},
	primaryclass = {astro-ph.CO}
}

@article{Dai:2013ikl,
	title        = {{Seeking Inflation Fossils in the Cosmic Microwave Background}},
	author       = {Dai, Liang and Jeong, Donghui and Kamionkowski, Marc},
	year         = 2013,
	journal      = {Phys. Rev. D},
	volume       = 87,
	number       = 10,
	pages        = 103006,
	doi          = {10.1103/PhysRevD.87.103006},
	eprint       = {1302.1868},
	archiveprefix = {arXiv},
	primaryclass = {astro-ph.CO}
}

@article{Chen:2006nt,
	title        = {{Observational signatures and non-Gaussianities of general single field inflation}},
	author       = {Chen, Xingang and Huang, Min-xin and Kachru, Shamit and Shiu, Gary},
	year         = 2007,
	journal      = {JCAP},
	volume       = {01},
	pages        = {002},
	doi          = {10.1088/1475-7516/2007/01/002},
	eprint       = {hep-th/0605045},
	archiveprefix = {arXiv},
	reportnumber = {SLAC-PUB-11840, MAD-TH-06-3, UFIFT-HEP-06-9, SU-ITP-06-12, CU-TP-1147}
}

@article{Acquaviva:2002ud,
	title        = {{Second order cosmological perturbations from inflation}},
	author       = {Acquaviva, Viviana and Bartolo, Nicola and Matarrese, Sabino and Riotto, Antonio},
	year         = 2003,
	journal      = {Nucl. Phys. B},
	volume       = 667,
	pages        = {119--148},
	doi          = {10.1016/S0550-3213(03)00550-9},
	eprint       = {astro-ph/0209156},
	archiveprefix = {arXiv},
	reportnumber = {DFPD-A-02-21}
}

@article{Dimastrogiovanni:2014ina,
	title        = {{Inflationary tensor fossils in large-scale structure}},
	author       = {Dimastrogiovanni, Emanuela and Fasiello, Matteo and Jeong, Donghui and Kamionkowski, Marc},
	year         = 2014,
	journal      = {JCAP},
	volume       = 12,
	pages        = {050},
	doi          = {10.1088/1475-7516/2014/12/050},
	eprint       = {1407.8204},
	archiveprefix = {arXiv},
	primaryclass = {astro-ph.CO}
}

@book{bialynicki2013quantum,
	title        = {Quantum electrodynamics},
	author       = {Bialynicki-Birula, Iwo and Bialynicka-Birula, Zofia},
	year         = 2013,
	publisher    = {Elsevier},
	volume       = 70
}

@article{Dai_2012,
	title        = {Total angular momentum waves for scalar, vector, and tensor fields},
	author       = {Liang Dai and Marc Kamionkowski and Donghui Jeong},
	year         = 2012,
	month        = {dec},
	journal      = {Physical Review D},
	publisher    = {American Physical Society ({APS})},
	volume       = 86,
	number       = 12,
	doi          = {10.1103/physrevd.86.125013},
	url          = {https://doi.org/10.1103%2Fphysrevd.86.125013}
}

@article{Bordin_2018,
   title={Light Particles with Spin in Inflation},
   volume={2018},
   ISSN={1475-7516},
   url={http://dx.doi.org/10.1088/1475-7516/2018/10/013},
   DOI={10.1088/1475-7516/2018/10/013},
   number={10},
   journal={Journal of Cosmology and Astroparticle Physics},
   publisher={IOP Publishing},
   author={Bordin, Lorenzo and Creminelli, Paolo and Khmelnitsky, Andrei and Senatore, Leonardo},
   year={2018},
   month=oct, pages={013–013} }

@article{Dimastrogiovanni:2022afr,
	title        = {{Primordial stochastic gravitational wave background anisotropies: in-in formalization and applications}},
	author       = {Dimastrogiovanni, Ema and Fasiello, Matteo and Pinol, Lucas},
	year         = 2022,
	journal      = {JCAP},
	volume       = {09},
	pages        = {031},
	doi          = {10.1088/1475-7516/2022/09/031},
	eprint       = {2203.17192},
	archiveprefix = {arXiv},
	primaryclass = {astro-ph.CO}
}

@article{Salopek:1990jq,
    author = "Salopek, D. S. and Bond, J. R.",
    title = "{Nonlinear Evolution of Long Wavelength Metric Fluctuations in Inflationary mMdels}",
    reportNumber = "FERMILAB-PUB-90-131-A",
    doi = "10.1103/PhysRevD.42.3936",
    journal = "Phys. Rev. D",
    volume = "42",
    pages = "3936--3962",
    year = "1990"
}

@article{Chen:2010xka,
    author = "Chen, Xingang",
    title = "{Primordial Non-Gaussianities from Inflation Models}",
    eprint = "1002.1416",
    archivePrefix = "arXiv",
    primaryClass = "astro-ph.CO",
    doi = "10.1155/2010/638979",
    journal = "Adv. Astron.",
    volume = "2010",
    pages = "638979",
    year = "2010"
}

\end{document}